\documentclass[prl]{article}

\usepackage{graphicx,psfrag} 
\usepackage{amsmath, amssymb}
\usepackage{colonequals}

\usepackage{geometry}
\usepackage{upgreek}

\begin{document}

\title{Tree Code for Collision Detection of Large Numbers of Particles\\{Application for the Breit-Wheeler Process\\$[$preprint$]$}}

\author{\bf O. Jansen, E. d'Humi\`eres, X. Ribeyre, \\ \bf S. Jequier, V.T. Tikhonchuk\\
	\small Univ. Bordeaux/CNRS/CEA, Centre Lasers Intenses et Applications\\
	\small oliver.jansen@celia.u-bordeaux.fr}

\date{\today}
\maketitle

\begin{abstract}
Collision detection of a large number $N$ of particles can be challenging. Directly testing $N$ particles for collision among each other leads to $N^2$ queries. Especially in scenarios, where fast, densely packed particles interact, challenges arise for classical methods like Particle-in-Cell or Monte-Carlo. Modern collision detection methods utilising bounding volume hierarchies are suitable to overcome these challenges and allow a detailed analysis of the interaction of large number of particles. This approach is applied to the analysis of the collision of two photon beams leading to the creation of electron-positron pairs.\\
\end{abstract}

\begin{center}
{\bf Keywords}\\
tree code; collision detection; QED; Breit-Wheeler process; pair creation; astronomy
\end{center}





\section{Introduction}
\label{Introduction}
Modelling a large number of particles often is a challenge in physics. Many-body problems are well known in astronomy, plasma physics, solid state physics and other disciplines. In astronomy a common way to overcome the challenge of simulating a many-body problem, like the movement of stars of one galaxy under each others gravitational force, is to use the Barnes-Hut (BH) method \cite{Tree}. In a BH simulation space is partitioned in an hierarchic octree structure. The tree branches grow towards successive smaller volumes of space in such way as to include at maximum one particle (star) in each leaf node, while still covering the entirety of the simulation domain. Only particles, that are in partitions close to each other, interact.\\ In plasma physics Particle-in-Cell (PIC) codes \cite{Buneman}\cite{Dawson} are regularly used. There, the simulated domain is structured in a mesh, with fields being defined on the nodes of that grid, while particles, combined in mono-energetic clouds called \textit{macro-particles} (MPs), can move freely in space. Vlasov- and Fokker-Planck codes as well are quite common in plasma physics. Tree codes like the BH in plasma physics are not very common, but exist \cite{PlasmaTreeCode}.\\
For many problems in solid-state physics Monte-Carlo (MC) \cite{MC} and Molecular Dynamics (MD) \cite{MD1}\cite{MD2} simulations are well suited.\\
Many of these simulations can require a lot of computational power. Dealing with collision detection in systems with many particles even further increases the burden on computational performance. In order to test an ensemble of $N$ particles on all possible collisions, one needs $N^2$ queries. This can easily become a computational effort too large to handle, which calls for different approaches and approximations in simulations. In PIC codes collisions usually are handled statistically and in a probabilistic fashion using MC methods or binary collision models \cite{BinColl}. In BH and MD simulations collisions usually are not treated directly, but via the potentials generated by the particles.\\
\newline
In other fields direct collision detections of moderate numbers of particles are very common. In airport short-term conflict alert systems compute possible collisions between aircrafts in order to avoid them. In video games and computer-generated imagery (CGI) over the years more and more objects interact in the same simulation environment with increasingly higher details. In the cases of airport traffic control and video games, performance is of utmost importance, since all calculations have to be done in real time. However, precision can not be allowed to suffer from the high demand on performance. In fact, modern methods are even more precise than their predecessors.\\
In the following we present a method, that combines some of the before mentioned techniques into one powerful and versatile tool for computing collisions of a very large number of particles. We present results from simulations, computed on a single workstation without parallelisation, including collisions of about $10^6$ particles in a single particle picture, up to $10^9$ particles in a slightly restrained way and arbitrary number of particles in a statistical approach. 
These results are obtained by using the scheme called Bounding Volume Tree Hierarchy Simulation for Interactions in Large Ensembles (TrI LEns), which adaptively partitions phase space in bounding volumes for efficiently pruning the problem of collision detection. 
Our main motivation is to develop a method, that allows an accurate single-particle description of collisions while dealing with a very large number of particles. It is important for us to obtain a tool, that can be used in different regimes with arbitrary large numbers of particles, even though the form of representation of particles and collision might change between regimes. This way we are able to investigate a transition from one regime to another and compare results from analytical predictions, simulation results with detailed descriptions of individual particles and statistical approximations.\\
\newline
In recent years, the collisions of photons with the creation of electron-positron pairs according to the Breit-Wheeler (BW) process \cite{BW} has become of particular interest in fundamental research and astrophysics. 
The BW process is the inverse process to the annihilation of an electron and a positron into two photons and it is therefore, the most basic process for the creation of matter from light. It is believed to be an important aspect in gamma ray burts \cite{Piran}, active galactic nuclei, black holes and other large-scale explosive phenomena \cite{Ruffini} in the universe. The BW process also is responsible for the $TeV$ cutoff in the photon energy spectrum of extra-galactic sources \cite{Nikishov}.\\
An experiment at the Stanford Linear Accelerator Center \cite{Burke}\cite{Bamber} was able to detect pairs created by the collision of photons, however, was unable to reach the regime of the linear BW process, in which exactly two photons create one electron and one positron. Ribeyre \textit{et al.} proposed a novel experiment \cite{Xavier} in order to investigate the BW process in laboratory conditions with a significant number of particles being created and detected with a minimum of noise sources. The actual collision of photons in this scheme takes place in a small volume in vacuum, far away from any sources of trident or Bethe-Heitler \cite{Bethe} pair creation. This experimental set-up could be implemented in soon upcoming facilities like Apollon \cite{Apollon} or ELI-NP \cite{ELI}. A more detailed analysis of the specific implementation of the experiment will be handled elsewhere. Even though, the dynamics of two photons in the BW process can easily be calculated, a large number of photons colliding in a small volume can become difficult to handle. In the following, we present a way of how this simulation code can be applied to investigations of the BW process. For the BW process the total cross-section, that defines the probability of two photons colliding, is given \cite{LaLi}\cite{QED} by
\begin{equation}
	\label{BWcross}
	\sigma_{\gamma \gamma} = \frac{\pi}{2}r_e^2(1-\beta^2)\left[ -2\beta(2-\beta^2) + (3-\beta^4)\ln\frac{1+\beta}{1-\beta}\right],
\end{equation}
with $\beta = \sqrt{1-1/s}$ and $s = E_{\gamma_1} E_{\gamma_2} (1-\cos\Phi_B)/(2m_e^2c^4)$, where the $E_{\gamma_i}$ are the photon energies, $c$ is the speed of light, $m_e$ is the rest mass of the electron, $\Phi_B$ is the angle under which collision occurs and $r_e$ is the classical electron radius. For reason of energy conservation, the BW process has the strict threshold of
\begin{equation}
	\label{eq.Cond}
	 E_{\gamma_1} E_{\gamma_2} \ge \frac{m_e^2c^4}{\frac{1}{2}(1-\cos\Phi_B)}.
\end{equation}

The denominator in the right hand side of equation \ref{eq.Cond} is caused by the centre-of-mass (CoM) movement of the photon pair and the electron-positron pair in case the two photons do not perfectly counter-propagate ($\Phi_B \neq 180^\circ$). In such a case, the CoM movement of the photons has to be carried over to the electron-positron pair. In the CoM frame the two created particles have the momenta
\begin{equation}
	\label{pTotal}
	p^\prime_{e,p} = \pm \sqrt{\frac{1}{4}\left[\left(\frac{E_{Total}}{c}\right)^2 - (\vec{p}_{Total})^2\right] - m_ec^2},
\end{equation}
where $E_{Total}$ and $\vec{p}_{Total}$ are the total energy and momentum of the two photons in the lab frame. One particle is then created with a relative velocity $\vec{\beta}^\prime = \vec{v}^\prime/c$ according to its momentum, while the other moves in the exact opposite direction. The exact direction of $\vec{\beta}^\prime$ is irrelevant in the CoM frame and in our simulations it is determined randomly. The relative velocity in the lab frame for each particle can then be calculated by Lorentz-transformation as
\begin{equation}
	\label{Lorentz}
	\vec{\beta} = \frac{\vec{\beta}_{CoM} + \vec{\beta}^\prime_\parallel + \vec{\beta}_\perp \sqrt{1-(\vec{\beta}_{CoM})^2}}{1 + \vec{\beta}^\prime \vec{\beta}_{CoM}},
\end{equation}
with $\vec{\beta}_{CoM}$ being the relative velocity of the CoM, $\vec{\beta}^\prime_\parallel$ the portion of $\vec{\beta}^\prime$, that is parallel to $\vec{\beta}_{CoM}$ and $\vec{\beta}^\prime_\perp = \vec{\beta}_{CoM} - \vec{\beta}^\prime_\parallel$.
We chose this example, because in the experimental scheme proposed by Ribeyre \textit{et al.} \cite{Xavier} two beams of photons collide in order to create electron-positron pairs. The two colliding beams consist of about $\sim10^{13}$ photons with a divergence of up to $20^\circ$. Simulating this huge number of particles alone would require a PIC simulation with very heavy macro-particles or a strictly statistical approach. One of our goals is to be able to simulate similar, but smaller set-ups in a very detailed fashion, which could be used as benchmarks. Furthermore, since a statistical approach is unavoidable for the full problem, we want the best statistical sampling possible, which meant increasing the number of macro-particles as much as possible.\\
\newline

In the following chapter \ref{BVHTC} we explain the main concept of \textit{Bounding Volumes Hierarchies} in detail in section \ref{BVH} and how they are implemented in the presented code. Since bounding volume hierarchies are the core concept of the TrI LEns code, in section \ref{MPs} we further discuss hierarchy traversal as a mean of collision detection and show, how particles are represented in these hierarchies.\\
Following this, chapter \ref{Application} deals with the application of the TrI LEns code for investigating the BW process. In section \ref{Performance} we compare the performance of the tree code with a corresponding mesh-based code using a scenario close to the experimental set-up, proposed by Ribeyre \textit{et al.} \cite{Xavier}. In the next section \ref{BW}, we provide simulation results specific to a BW inspired, possible experimental set-up, where we compare simulation results with theoretical predictions. We also include predictions, based on simulation results, that surpass the scope of currently available theories.\\
We end this publication with a summary of our results, including perspectives for further development and applications of the code.
\section{Bounding Volume Hierarchy and Tree Code}
\label{BVHTC}
In this section, we present the basic structure and features of the TrI LEns code. Since it is developed for the kinematics of a large number of particles, we use particle-in-cell (PIC) codes as a reference.\\

\subsection{Bounding Volume Hierarchy}\hfill\\
\label{BVH}
The first, obvious difference between a tree-  and a PIC code is the lack of a mesh in tree codes. Even though the tree code does not have a mesh, it still features the main benefit of a grid-based simulation in regard to collision detection. This feature is the reduction of the number of particles that actually are checked for collision, by only regarding particles that are spatially close to each other. In most mesh-based particle simulations, particles are grouped spatially in cells. These cells exist independently of particles and are often linked to their respective next neighbours in order to support the exchange of particles moving between cells. A tree code, however, is structured by dividing of particles into groups. The groups can be defined by particles, that are spatially close to each other (like in the BH method), but any other condition can be used in order to define groups. The groups and their partitions define nodes in a tree hierarchy, beginning with a single node containing all particles at the lowest level of the hierarchy. If the grouping condition is spatial vicinity then the nodes are quite similar to cells in a mesh-based code. The main difference, however, is that the nodes are very dependent on the contained particles. This leads to a very adaptive cell-structure comparable to an adaptive mesh \cite{AdaptMesh} in PIC-codes.\\

\begin{figure}
	\centering
	\includegraphics[width = 0.6\textwidth, angle = 0]
	{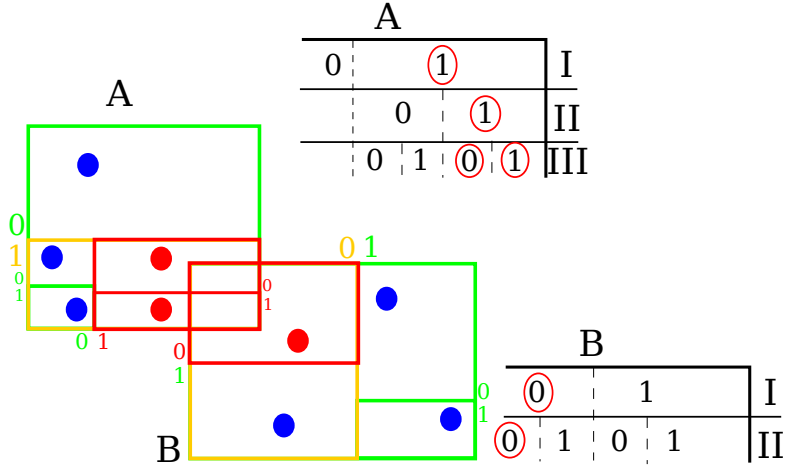}
	\caption{\textit{Two hierarchies of BVs overlapping with each other. {\normalfont Left}: the bounding volumes of each particle group and their partitions. Green partitions did not intersect with partitions of the other group, yellow ones did, but only on a low level and not on the highest one. The red partitions correspond to leaf nodes that intersect and include the only particles that have the potential for a collision during this time step. {\normalfont Right}: Hierarchy structure of the BVs and their partitions divided into their respective levels. Red circles denote intersections.}}
	\label{Hierarchies}
\end{figure}

The nodes in the tree hierarchy of the TrI LEns code are defined by \textit{Bounding Volumes} (BVs) iterated in BV hierarchies \cite{Coll}. Usually BVs are closed, simple volumes, that entirely encapsulate a group of particles or polygons. Instead of testing all members of one group of particles with all the members of another group, one can first test both corresponding BVs. If there is no overlap between the BVs, then there is no collision between the contained groups.
Each BV, which we will refer to in the following as level $1$ BV, can be divided into a set of smaller BVs (of higher levels) in a way, that the union of the sub-BVs contains all the particles of the original BV. The same method of quick and rough collision checks can be done using sub-BVs in case the two original BVs overlap. This method of subdividing level $i$ BVs into level $i+1$ BVs defines a hierarchy structure for each group of particles. Depending on the algorithm used to traverse the hierarchies and the shape of the BVs, those rough collision checks can be very swift, possibly avoiding any vector product and square root, while reducing the number of accurate, more expensive collision tests between real particles significantly. Another benefit of using BV hierarchy is the fact that it is very adaptive, since it creates its "cells" depending on the particle distribution in a balanced way. Also, empty space, which in a mesh-based code still would be covered with cells, most often can be easily ignored.\\
This method of grouping objects into hierarchies of BVs in order to speed-up collision detection is well known in video games and computer-generated imagery (CGI). In both fields, all virtual objects consist of a number of elemental polygons, often triangles. As an example taken from these fields, let us assume a virtual car, consisting of polygons, which will serve as a group of particles. This object corresponds to a hierarchy, which incorporates its polygons, while other objects (for instance a road) have their own hierarchies. Collision detection between polygons of the same object are not required (the car does not collide with itself), which already reduces the number of collision tests significantly. If two hierarchies overlap, collision tests between all polygons of both hierarchies are rarely necessary. Instead, after traversing through both hierarchies using swift overlap checks for the bounding volumes and the sub-volumes, one usually finds, that only collision tests between a small subset of polygons of both hierarchies have to be done. In our example, usually it is only necessary to test the tires of the car against the polygons of one section of the road, ignoring the majority of the polygons of both car and road. An example of using BV hierarchies for the collision of two groups of particles is illustrated in Figure \ref{Hierarchies}. Here, particle group $A$, consisting of five particles, is partitioned in a three-level hierarchy, while group $B$, having four particles, has only two-levels. Unlike the BH simulation space has not been divided in an octree (or rather quadtree, since it is just a 2D example) structure. In the scheme, that is presented here, no node, that is entirely empty, exists in the tree hierarchies. A main difference to commonly used BVs is, that we added the distance, each particle of one group travels during one time step, to the size of their BVs. This way, we are able to pre-determine collisions instead of looking for collisions, that happened during a recently past time step. \\
Before describing the hierarchy structure used in the TrI LEns code, we would like to clarify on the nomenclature used in this paper. A \textit{child node} is a node in a tree hierarchy of level $2$ or higher, usually with a reference to its directly related \textit{parent node}, one level below it, i.e. "The two level $i$ nodes $A$ and $B$ are the two children of the level $i-1$ parent node $C$". Furthermore, \textit{leaf nodes} are nodes that do not have children related to them. Let us denote the partitions or child nodes of a hierarchy $A$ in the following way: on each level, the partitions of the previous level are each divided into two new partitions, corresponding to two child nodes each, called $0$ and $1$. In order to denote, which node has been partitioned, we will write "$j$ of $i$" or "$ij$", where $j$ is either the node $0$ or $1$ and $i$ is the identifier of the parent node. As an example, the child node $1$ (level $2$) of the parent node $0$ (level $1$) of hierarchy $A$, therefore can be referred to as $A01$. As can be seen in the figure \ref{Hierarchies}, node $1$ of $A$ and $0$ of $B$ are the first ones at level $1$ to overlap (yellow boundaries) while the two other nodes ($0$ of $A$ and $0$ of $B$) are of no further concern. In case of $B$ the traversal through the hierarchy ends with the child node $B00$, which is the only node on level $2$ that intersects with any BV of hierarchy $A$. The only leaf nodes of $A$, that intersect with $B$ are $A110$ and $A111$. This reduces the question, which of the total of nine particles collide ($81$ queries) to the question, which of the two particles in the nodes $A111$ and $A110$ might collide with the particle in the node $B00$ (two queries). Note, that the terms "BV" and "node" are often used interchangeable. The reason for this is, that every node corresponds to exactly one BV and whether it is treated as a "node" or a "BV" depends mainly on the context. Also, the tree hierarchy of $A$ is not fully occupied, which means, that not all of the leaf nodes are of the same level. This reflects the main structure of the presented code, which features a tree structure, that also is not necessary fully occupied. The reason for this is the attempt to develop a dynamic structure, that is updated every time step in contrast to other methods, in which the tree structure has to be entirely re-calculated. However, as a trade-off one might have to remove or add nodes, if particles are destroyed, created or simply change their momentum (see below). The TrI LEns code, unlike the BH approach does not include any node that covers empty space.\\

Since gathering particles into groups is essential for the scheme proposed here, one has to find reasonable conditions for grouping the particles into hierarchies. The smaller a BV is, the less likely it is to overlap with others. It is, therefore, desirable to find the smallest, possible BV or sub-BV for each group of particles. Also, particles with the same velocity and direction of movement do not collide, making it unnecessary to perform collision tests within one such group. For these reasons and because we increase the size of the BVs by the distance the included particle travels each time step, the most practical course of action is to group particles that move parallel to each other. This means, that we do not just divide the space into BVs, but the phase space. As a threshold for the parallel movement of particles, we chose eq. (\ref{eq.Cond}). This strict condition is easily computed for each particle at the beginning of the simulation. However, other conditions are conceivable, too. With more or less parallel moving particles it is possible to choose the size of each BV in a way that no particle will ever leave their respective BV while still having BVs that are not much larger than unmodified ones for stationary particles.\\
\newline

The BVs implemented in this code are axis-aligned minimum bounding-boxes (AABB). Therefore, collision tests between BVs become quite simple. For two BVs $1$ and $2$ defined by two opposing corners $\vec{C}_{low, i} = (lx_i, ly_i, lz_i)$ and $\vec{C}_{high, i} = (hx_i, hy_i, hz_i)$, where $i=1,2$ stand for BV $1$ and BV $2$ respectively and $l_{k_i} \le h_{k_i}$ for $k=x,y,z$ the collision test can be reduced to simple comparisons of the positions of the corners. By testing (in pseudo-code)
\begin{align}
  \label{CollDet}
	if (&lx_1 < hx_2 \text{ AND } ly_1 < hy_2 \text{ AND } lz_1 < lz_2 \text{ AND }\\
	\notag
	&hx_1 > lx_2 \text{ AND } hy_1 > ly_2 \text{ AND } hz_1 > lz_2)
\end{align}
\begin{figure}
	\centering
	\includegraphics[width = 0.6\textwidth, angle = 0]{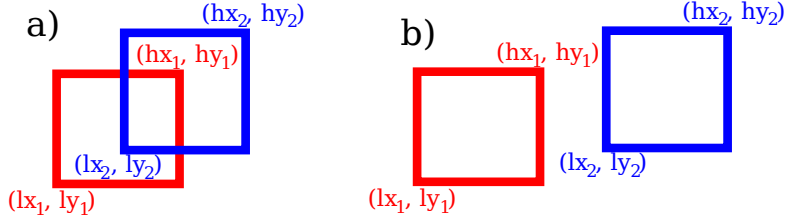}
	\caption{\textit{a) Two boxes, determined by two opposing corners each, overlap. Equation \ref{CollDet} is fulfilled. b) The two boxes do not overlap, because just one condition for overlap $hx_1 > lx_2$ is not fulfilled.}}
	\label{Overlap}
\end{figure}
it is possible to entirely avoid square roots and float point multiplications. An example for overlap can be seen in Figure \ref{Overlap}. Please keep in mind, that "minimum bounding box" does refer to the phase space partitioning and therefore, leads to larger BVs as one might expect from regular minimum bounding boxes. The shape and size of each BV is not adjusted during the simulation for performance reasons. Therefore, they have to be calculated during initialisation in a way to ensure, that all particles stay inside their BVs during the entire simulation. This is possible in the case that collisions do not result in change of momentum as it is the case with photons colliding and annihilating. If collisions result in change of momentum or if particles are added during simulation, one has to create a hierarchy every time step analogous to the BH simulation or possibly have to perform significant updates to existing hierarchies.\\
The bounding boxes in general are not cubes, nor do they cover the entire simulation space. The exact shape of each box is calculated after particles have been assigned to it and it is trimmed down to the smallest size suitable. Whenever not specified otherwise, please note that a box shape is used. It should be noted, that AABBs suffer from certain scenarios, as for instance two particle beams colliding under an angle, that is half the angle between the axes. However, we decided against alternatives like oriented bounding boxes (OBB), because of the computational effort and complexity of collision tests between OBBs. A future upgrade aims to include bounding spheres.\\
\newline
The hierarchies for the groups of particles are represented with linear k-dimensional (k-d) trees, where $k=2$. The position for each cut, creating the next level of the k-d tree, is calculated by distributing the contained particles as evenly as possible. The direction, in which a cut occurs, is cycled through after each cut. The reason for this choice over an octree, as being used in the BH simulation, is the fact that efficiently reducing the number of particles per BV and having a small number of BVs is essential in our case. Our choice of partitioning usually leads to non-cubic bounding boxes, which have the advantage to fit their encapsulated particles more tightly than strictly cubic ones. This is preferable, since unnecessarily large BVs lead to additional overlaps with other BVs, that do not lead to actual particle collisions.\\

\subsection{Hierarchy traversal and macro-particles}
\label{MPs}
\begin{figure}
	\centering
	\includegraphics[width = 0.6\textwidth, angle = 0]
	{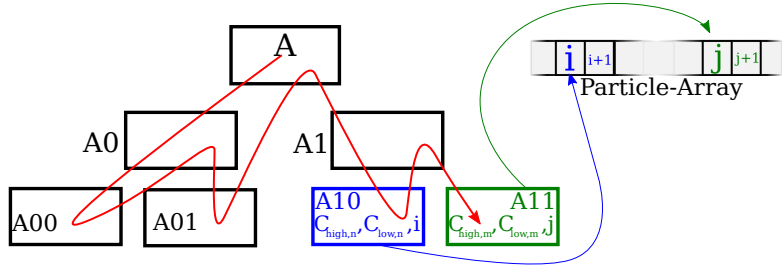}
	\caption{\textit{Traversal order in a depth-first tree traversal and link between leaf nodes and the particle array.}}
	\label{ListTraverse}
\end{figure}

In order to find colliding BVs in two or more hierarchies, an efficient traverse method is needed. Without a detailed comparison of possible methods, we chose a depth-first descent algorithm. We decided on depth-first descent, since we did not expect any particle groups of significant spatially different dimensions to occur. Those scenarios (for instance, a small object flying through a large building) usually benefit from more sophisticated descent rules. However, those rules come at the cost of greater complexity of the source code and higher computation cost in simple scenarios, like the ones, we are expecting. 
Although, less memory-intense methods are possible, linked lists are used both for the hierarchy traversal and for connecting the hierarchies with the particles. Possible collisions are chronologically sorted in order to avoid causality problems. Figure \ref{ListTraverse} illustrates the depth-first descent of hierarchy $A$. The descent method chooses the shortest way to the leaf node with the lowest running index, that is allowed by the collision detection condition (\ref{CollDet}). From there, in a similar fashion collision checks for traversal towards the next leafs are conducted. Please note, that should for instance the collision check for the node $A0$ be negative, its children $A00$ and $A01$ will be skipped. Each node contains information about the position of its two defining corners $\vec{C}_{low, i}$ and $\vec{C}_{high, i}$. In addition to that, leaf nodes contain a link to the particle array towards the position of the first particle, that is contained in the leaf, as well as the number of particles that the leaf holds. The particle array is sorted according to the affiliation of the particles to their leaf nodes.

In order to support large particle numbers and for a more intuitive interface to PIC-codes, we implemented a particle description with macro-particles (MPs). MPs can be adjusted to represent exactly one real particle. In this regime they are nothing but the smallest possible BV in each hierarchy and the code treats each particle individually. However, in order to increase the number of particles with a comparably small impact on performance MPs, representing more than one real particle, can be used. In order to satisfy an individual particle (IP) representation in this regime, MPs are "split" into swarms of real particles, should two MPs collide. These swarms consist of particles with the exact same momentum, but different positions. This corresponds somewhat to the discretisation of MPs in momentum space in PIC-codes. In order to be able to handle collision detection of two heavy MPs the real particles of each MP are embedded into an octree BV-hierarchy again. The use of an octree was motivated by the real particles of one MP being positioned as homogeneously as possible over the volume of the MP. Also, the hierarchies for the real-particles have to be created during each collision. A fast and simple partitioning method seemed reasonable.\\

As a mean to handle arbitrarily high particle numbers, it is also possible with the TrI LEns code to leave the IP representation behind and treat the MPs as continuous clouds of particles corresponding to MPs in PIC simulations. In this statistical (ST) regime collision rates between two MPs are calculated instead of collisions between real particles. The rate of collision $\dot{n}(t)$ between two such clouds $I$ and $II$ can be calculated as
\begin{equation}
	\label{rate}
	\dot{n}(t) = n_I(t) n_{II}(t) <\sigma v>,
\end{equation}
where $n_I(t)$ and $n_{II}(t)$ are the number densities of the clouds and 
\begin{equation}
	\label{reactivity}
	<\sigma v> = \int\int d\vec{v}_I d\vec{v}_{II} \sigma_{1,2}(v) v f_I(\vec{v}_I)f_{II}(\vec{v}_{II}),
\end{equation}
is the average reactivity with the cloud velocities $\vec{v}_i$, a cross section $\sigma_{1,2}(v)$ and the distribution $f_i(\vec{v}_i)$ of $\vec{v}_i$, $i=I,II$\cite{LaLi}\cite{InFusion}. For two photon clouds ($|\vec{v}_i|=c$, $\vec{v}_I\cdot \vec{v}_{II} = \cos\phi$ and $v = c(1-\cos\phi$) and the BW cross section (eq. (\ref{BWcross})) the reactivity simplifies to
\begin{equation}
	<\sigma v>_{BW} = 2\pi \sigma_{\gamma\gamma} c.
\end{equation}
With this, we can write the pair production rate
\begin{equation}
  \label{DGL}
	\dot{n}(t) = 2\pi n_I(t) n_{II}(t) \sigma_{\gamma\gamma} c.
\end{equation}
and the conservation of particles
\begin{align}
\notag
	n_I(t) &= n_I(0) - n(t)\\
	n_{II}(t) &= n_{II}(0) - n(t).\\
\end{align}
By introducing the following notations
\begin{equation}
	n_0 \colonequals n_I(0) + n_{II}(0)
\end{equation}
and
\begin{equation}
	\alpha \colonequals \frac{n_I(0)}{n_0}
\end{equation}
we can integrate (\ref{DGL})
\begin{equation}
	\label{DGLint}
	\int_0^{n(\tau)} \frac{dn}{\left(\alpha n_0 - n\right)\left( (1-\alpha)n_0 - n\right)} = 2\pi\sigma_{\gamma\gamma} c \tau.
\end{equation}
In case of $\alpha = 1/2$, (\ref{DGLint}) simplifies into
\begin{equation}
	\int_0^{n_{1/2}(\tau)} \frac{dn_{1/2}}{\left(\frac{n_0}{2} - n_{1/2}\right)^2} = 2\pi\sigma_{\gamma\gamma} c \tau,
\end{equation}
which gives
\begin{equation}
	n_{1/2}(\tau) = \frac{2\pi \left(\frac{n_0}{2}\right)^2\sigma_{\gamma\gamma}c\tau}{1+ \pi n_0 \sigma_{\gamma\gamma}c\tau}.
\end{equation}
For any $\alpha \neq 1/2$ and $0<\alpha<1$ (\ref{DGLint}) can be solved as
\begin{equation}
	n_\alpha(t) = \frac{(\alpha-1) \alpha n_0 \left(1 - e^{(1-\cos\theta) n_0(2\alpha-1) \sigma_{\gamma\gamma}c\tau}\right)}{\alpha e^{(1-\cos\theta) n_0(2\alpha-1) \sigma_{\gamma\gamma}c\tau} + \alpha - 1}.
\end{equation}
The solutions for $n_{1/2}$ and $n_\alpha$ take into account the reduction of possible collision partners during the interaction of two clouds due to the annihilation of photons. To approximate the exact volume of the two clouds that overlap, the same octree structure that is used in the IP representation, can be utilised.\\
We therefore can work within three different regimes with the TrI LEns code: The first regime represents each particle individually (similar to macro-particles with a number weight of $1$). The second regime combines several particles into a swarm of co-moving particles with the same momentum. During collisions, all of those particles are handled individually, which is why we refer to this as the \emph{IP regime}. The third regime is closest to the one used in PIC-simulations, since it treats particles as continuous clouds. It is a statistical picture (referred to as the \emph{ST regime}) that is very useful for very high particle numbers. Even though the description of the particles changes quite significantly, results and most importantly, the transitions between these artificial, purely numerical regimes match analytical results quite well (see section \ref{BW}).\\

Since collisions in the case of the Breit-Wheeler process create new particles and the number of macro-particle created can become extremely high, we also implemented a merging method for the created particles. We used the merging algorithm presented by Vranic \textit{et. al} \cite{Merge}.\\

\section{Application of the code}
\label{Application}
\subsection{Analysis}
\label{Performance}
\begin{figure}
	\centering
	\includegraphics[width = 0.6\textwidth, angle = 0]{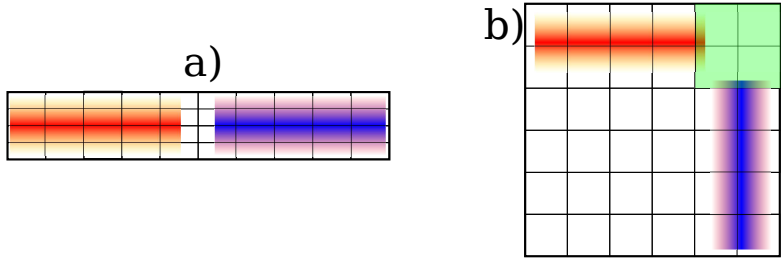}
	\caption{\textit{Two beams (red and blue) are about to collide with each other. The black boxes denote a 2D simulation mesh, belonging to PIC simulations. a) Well suited scenario for a PIC simulation. Total computation effort is well spread over most cells. b) Problematic geometry, that leads to larger cell sizes and all collisions taking place in just the four marked cells.}}
	\label{PICcomp}
\end{figure}

In this section, we elaborate on the reasons for using a tree code, while directly analysing its performance. The reason for using a tree code-like structure in the following scenario becomes apparent, if one considers using a PIC-code instead. One might consider a scenario, in which two beams of particles counter-propagate on the same axis with only a small divergence. In such a case, a PIC code would be well suited in order to describe the collision of both beams, since the simulation domain and with it the mesh more-or-less is given just by the size of both beams (see Figure \ref{PICcomp} a)). The resolution can be chosen very fine in order to make sure, that not too many particles are inside the same cell at the same time. However, already the first problem arises: very small cells lead to small time steps, since otherwise particles would jump cells and avoid collisions that otherwise might have occurred. Even worse than this would be a scenario, in which both beams do not move anti-parallel, but instead orthogonal to each other (see Figure \ref{PICcomp} b)). Alternatively, one can consider a scenario, in which the mesh is given, in order to cover certain important structures for instance, and in which two collimated beams would be to collide. In both cases the mesh will cover empty space while at the same time most likely will not have a high enough resolution as to keep the number of particles per cell small enough for the computational effort to be manageable. If the simulation domain and the mesh cannot be pruned very precisely on a problem, which requires detailed knowledge about the particle distribution, then PIC codes are prone to suffer from performance problems due to large number of particles per cell in certain scenarios. Of course, one can implement merging algorithms for macro-particles, but this always comes at the cost of deteriorating the statistical sampling of the phase space distribution given by the macro-particles. It should be mentioned, that there exist PIC codes, that utilise adaptive meshes \cite{AdaptMesh}, which proved to be very useful for thin target interactions for example. However, their complexity and certain related challenges make them difficult to handle and implement properly. Therefore, it is not advisable to use them unless necessary. In particular in the case of the collision of two beams of particles, where no external or background fields are involved, a mesh does not seem to be entirely necessary.\\

\begin{figure}
	\centering
	\includegraphics[width = 0.6\textwidth, angle = 0]{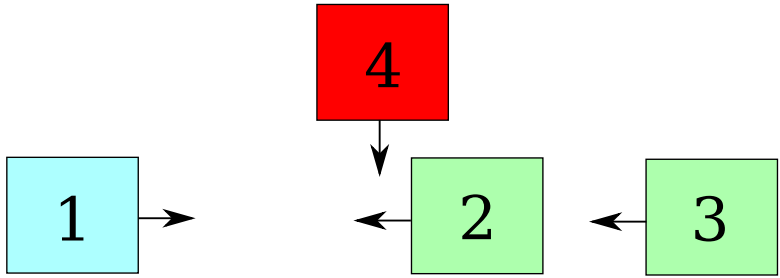}
	\caption{\textit{Particle $A=1$ is going to collide with particle $B_1=2$ first and then with particle $B_2=3$. Particle $1$ and $4$ are not going to collide. However, before particle $1$ and $2$ collide, $2$ may collide with $4$, which could have an impact on the collision between $1$ and $2$.}}
	\label{ChronOrder}
\end{figure}

\begin{figure}
\begin{minipage}{0.45\textwidth}
	\centering
	\includegraphics[width = \textwidth, angle = 0]{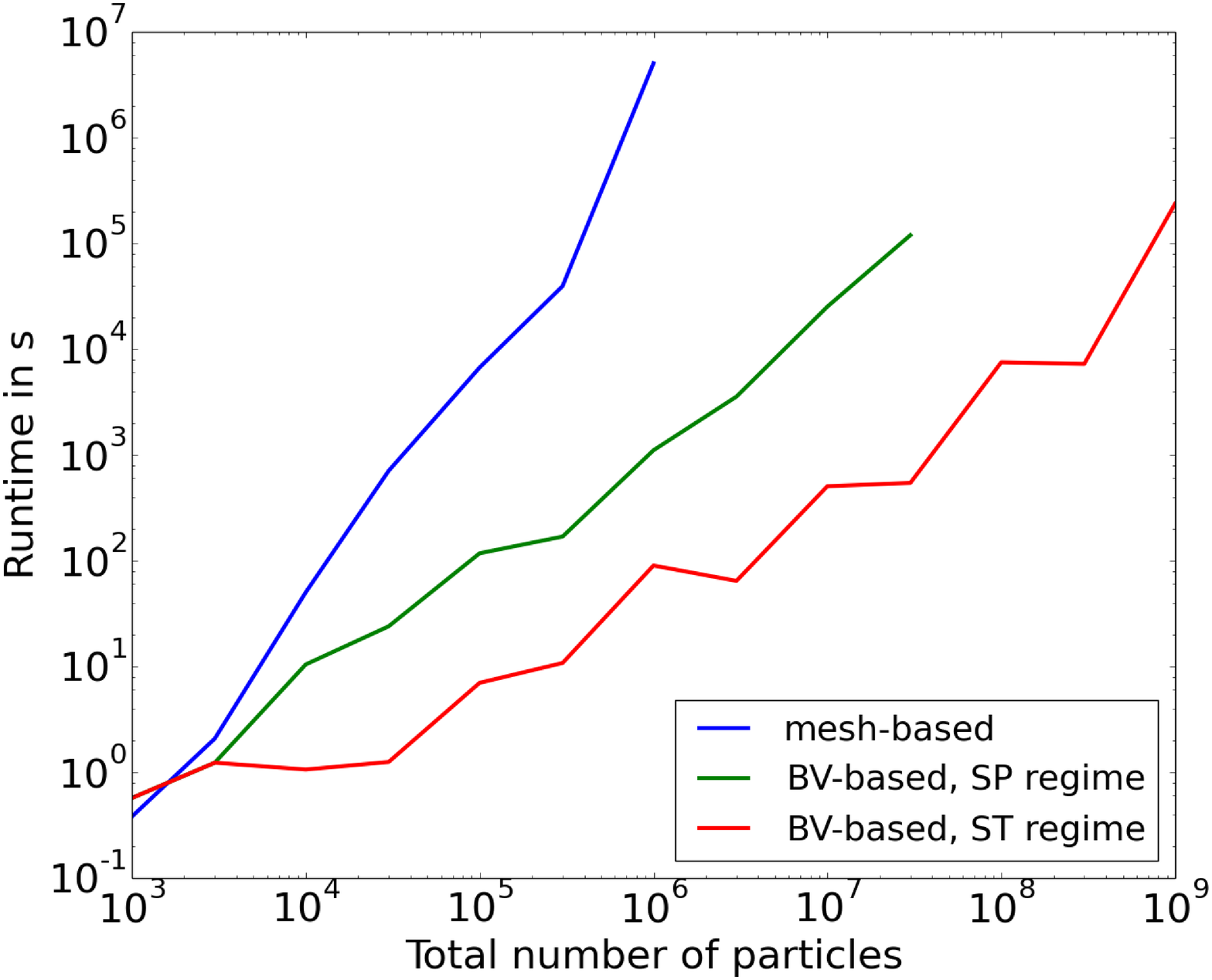}
\end{minipage}
\hspace{0.05\textwidth}
\begin{minipage}{0.45\textwidth}
	\centering
	\includegraphics[width = \textwidth, angle = 0]{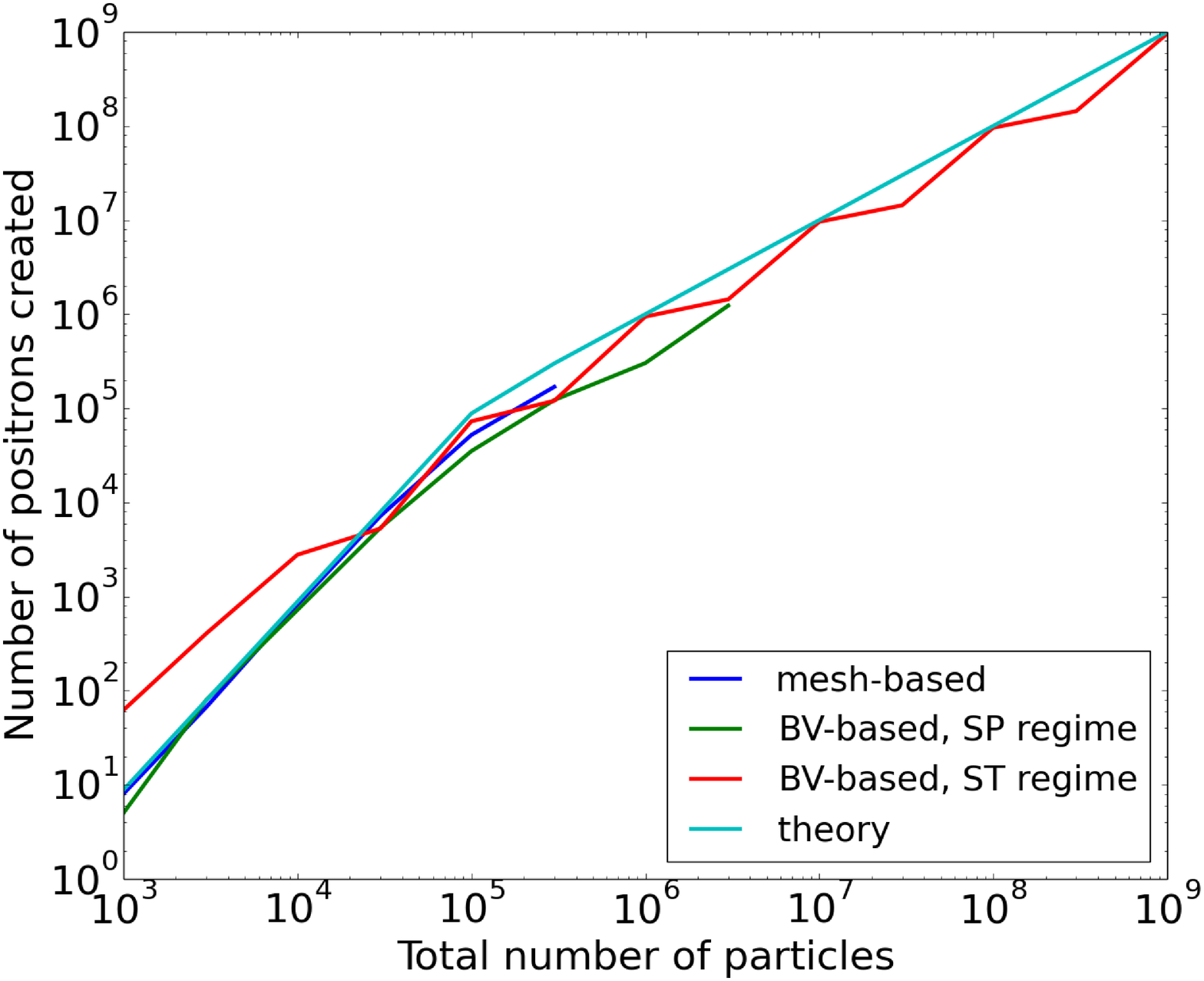}
\end{minipage}
\caption{\textit{Plotted are simulation results for mesh-based code (blue), the tree code in the IP regime (green) and the ST regime (red). Cyan in the right-hand plot denotes the theoretical predictions. \emph{Left:} Comparison of run times. \emph{Right:} Comparison of the number of pairs created.}}
\label{GridVsBVtree}
\end{figure}

In order to test the performance of the TrI LEns code, we compared it with mesh-based simulations. For that reason, we developed an alternative version, that operates entirely analogously, but with the use of a mesh instead of BV hierarchies. The time step in the mesh-based code was adjusted, so that passing through several cells at once was impossible for the particles. The reason for this is the fact, that collisions only occurred between particles in the same or neighbouring cells. The cell size was adjusted to accommodate the increasing number of particles from $100$ up to $10^6$ cells. Larger cells allowed larger time steps, but larger number of particles per cell could easily dominate the performance. For the tree code, no adjustments were necessary, which also means, that no initial knowledge about the problem was required. In Figure \ref{GridVsBVtree} the difference between the run times of the mesh-based version of the code and the tree code is shown. We considered two beams of particles collided with each other under an angle of $90^\circ$. The diameter of the beams was $d = 10^{-12}$m, the photons were given a Lorentz-factor of $\gamma = E_\gamma/(m_ec^2) = 4$, where $E_\gamma$ is the energy of the photon. For the mesh-based simulation, we defined a mesh, covering a box of side length $L=3$nm. The initial positions of the beams were at $x_1=-5d, y_1=0, z_1=0$ and $x_2=0, y_2=-5d, z_2=0$. For the mesh-based simulation the simulation domain was well tailored to the problem, representing a significant amount of initial knowledge about the problem. Its size was chosen as $L_x = 7d, L_y = 7d, L_z = 4d$. The size of the beams is very small compared to realistic parameters. The choice was made, in order to be able to compare the IP and the ST regime directly with each other and theoretical prediction. A simulation featuring two beams of a more realistic size in the range of $1-10\upmu\text{m}$ would require a very high number of photons in order to achieve a statistical relevant number of collisions. However, for purely numerical reasons we investigated a scenario, which can easily be simulated in either particle regime with reasonably short simulation durations. Section \ref{BW} features simulations with significantly more realistic parameters.\\
The time step was chosen automatically within the simulation in order to match the cell size and the velocity of the particles, which moved with a speed of $v=c$. Therefore, the calculated time step changed with changing cell size. For the tree code, the calculated, optimal time step was $c\Delta t_t = 3.93 d$. Here, the time step is calculated according to the size of the smallest BV, not including the macro-particles, in order to minimize the number of collisions each BV experiences during one time step. The reason for this restriction is the difficulty to order the collisions chronologically. It is not too complicated to determine all the possible collision partners $B_i$ for a  particle $A$ and order them chronologically. However, making sure, that any specific particles $B_k$ might not have collisions before colliding with $A$ is challenging at best. The most common approach of handling collisions is to pick a particle $A$ and then cycle through all possible interaction partners. After having created a list of all collision partners for $A$, one can sort the collisions chronologically. Unfortunately, without knowledge of collisions involving each particle $B_i$ on that list one can not be sure, that a collision between a specific $B_k$ and another particle $C$ occurred prior to $B_k$ collision with $A$ (compare with Figure \ref{ChronOrder}). The only way of avoiding this problem entirely is to save all collisions, that happened during one time step, in order to sort them chronologically. However, in the worst case scenario, that would lead to a list of $N^2$ items, which is infeasible for large $N$. Using small time steps, which reduce the number of multiple collisions for each particle, mitigates this problem. However, if one is not too concerned about the chronological order, the time step can be chosen even arbitrarily higher, which might speed up the simulations. In TrI-LEns, the choice can be done by the user on a case-to-case basis. As can be seen from Figure \ref{GridVsBVtree} on the left, the tree code had a performance one order of magnitude higher than the mesh based code. Only at very small numbers of particles did the mesh-based code show a superior performance, due to the large over-head of creating a tree-hierarchy. For the mesh-based code, $10^5$ particles was about roughly the maximum, that could be achieved with run times of less than a day. The tree code, however, with its  significantly better performance, was able to handle much more particles as can be seen in Figure \ref{GridVsBVtree} left. The blue curve denotes the mesh-based simulation, the green curve the corresponding tree-code simulation in the individual particle (IP) regime, in the special case where each macro-particle (MP) contains exactly one photon. Starting from $10^5$ particles, the tree-code achieved at least one order of magnitude shorter run times. At about the same run time, $10^7$ particles were possible to handle. The red curve shows simulations, in which groups of $10$, $100$ or $1000$ photons were treated as "clouds" as described in section \ref{MPs} as the more statistical ST regime. With this slight restriction on the momentum distribution of photons, the performance can be increased by at least one order of magnitude more. By carefully choosing the number of particles per MP, one can mitigate any possibly occurring statistical error. In Figure \ref{GridVsBVtree} on the right the actual simulation results of the number of positrons created are shown for the three cases. A good agreement between all cases is found. The only major deviation is given by the ST simulations for a small total number of particles. For only a few particles a statistical approach does not seem to be very accurate. Please note, that the point of discontinuity in the theoretical curve comes from the fact, that the number of created positrons is limited by the number of photons. At the discontinuity, the theory without that limit would predict more created pairs than creating photon pairs. Since the theoretical model does not take into account the exact space- and time dependent distribution of photons during the collision at very high densities the annihilation of all particles can be predicted. However, this is an unlikely event, since while collisions occur the density of both beams reduces and the likeliness, that at least some photons might pass through without colliding, increases. In such extreme scenarios, the analytical model, therefore, predicts too many collisions.

\begin{figure}
\begin{minipage}[t]{0.3\textwidth}
	\centering
	\includegraphics[width = \textwidth, angle = 0]{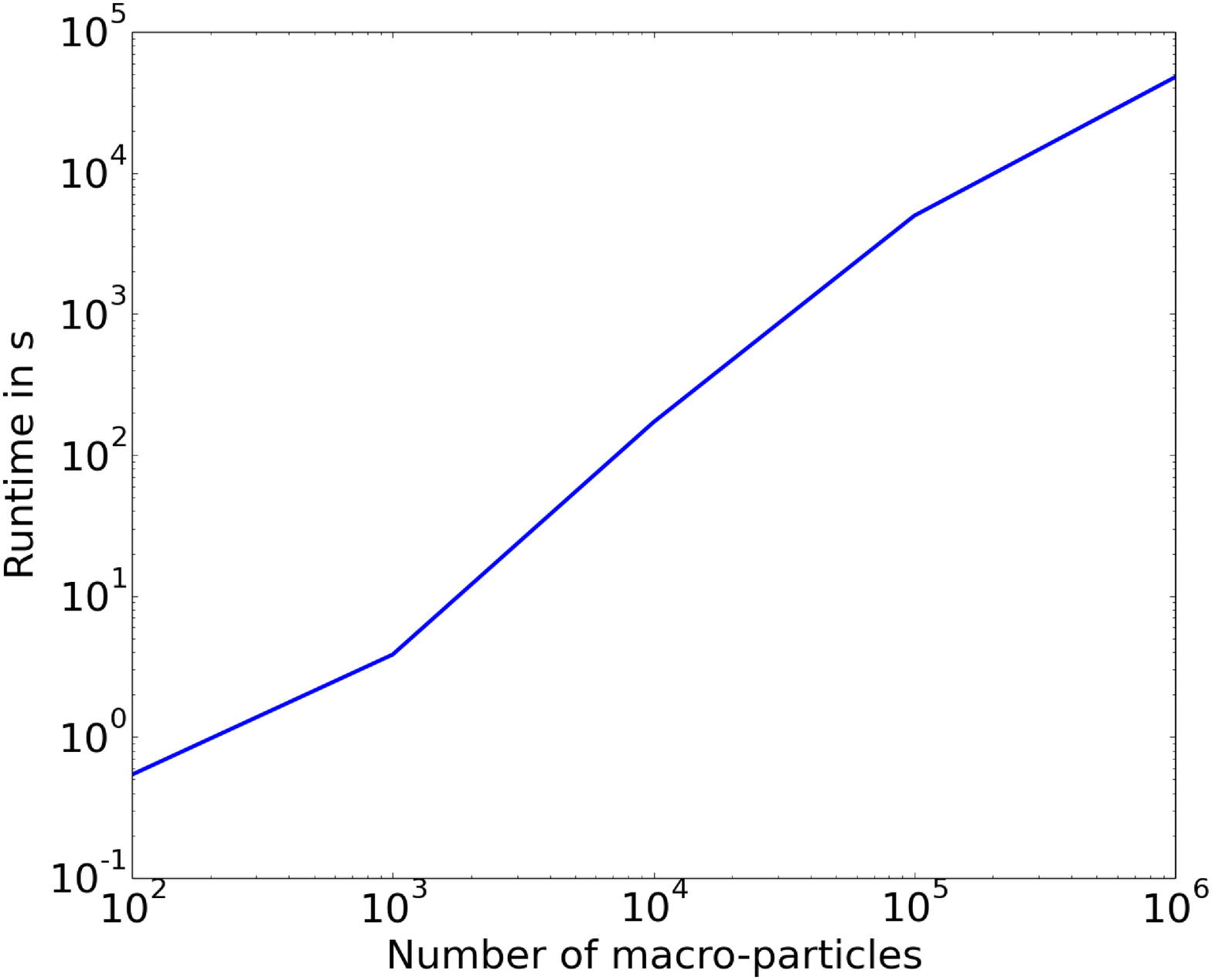}
\end{minipage}
\hspace{0.02\textwidth}
\begin{minipage}[t]{0.35\textwidth}
	\centering
	\includegraphics[width = \textwidth, angle = 0]{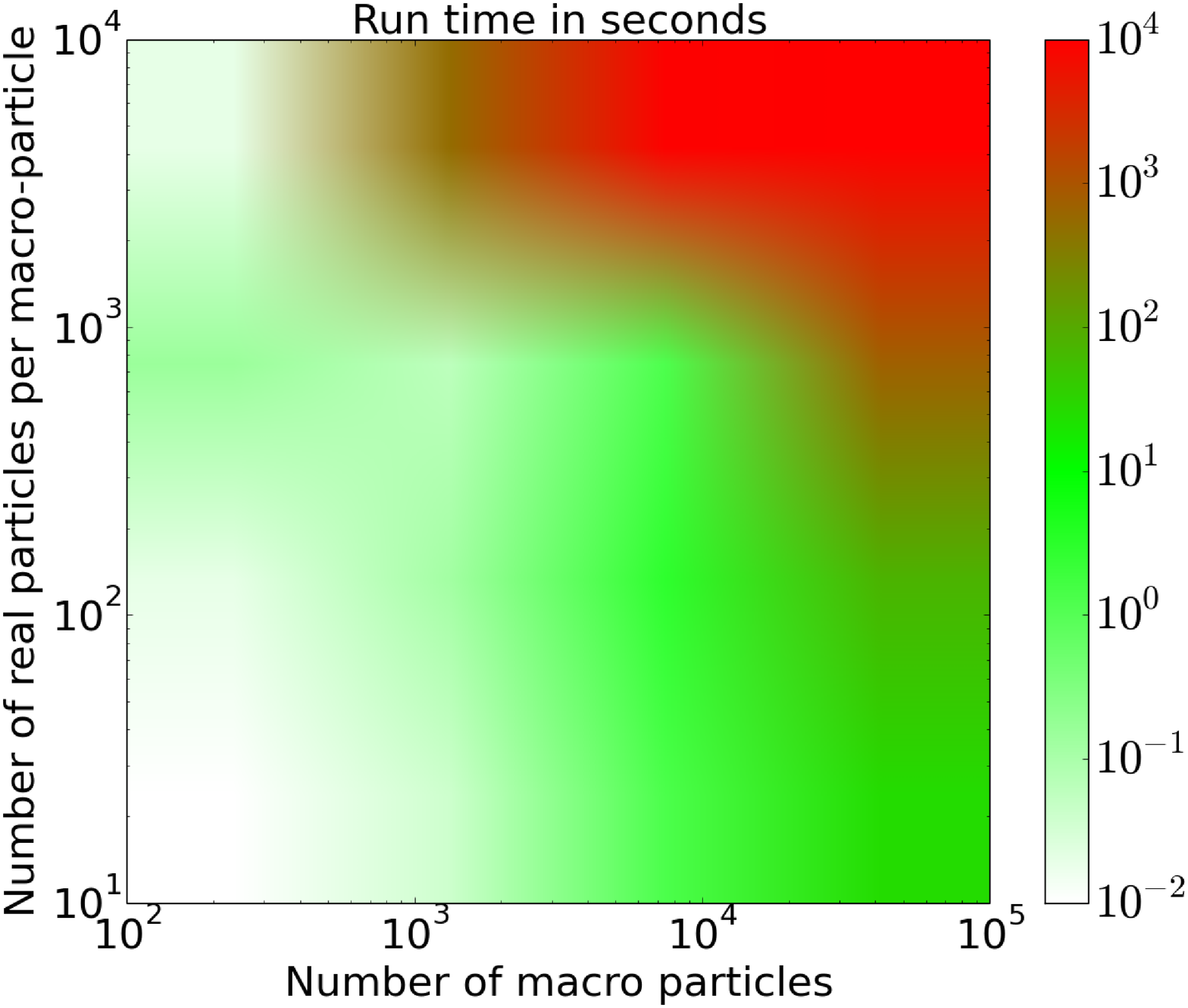}
\end{minipage}
\hspace{0.02\textwidth}
\begin{minipage}[t]{0.3\textwidth}
	\centering
	\includegraphics[width = \textwidth, angle = 0]{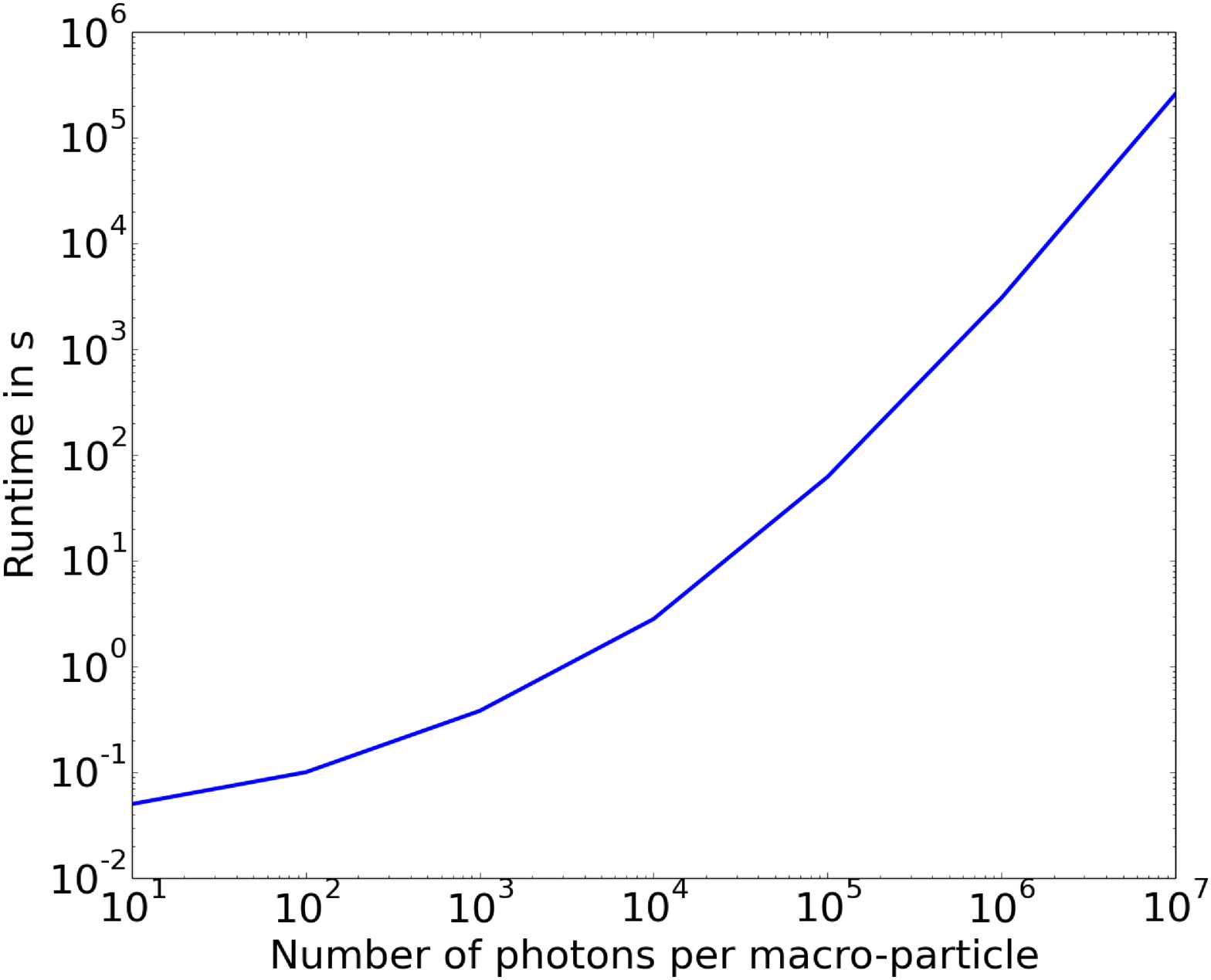}
\end{minipage}
\caption{\textit{\emph{Left: }Run time as a function of the number of macro-particles with constant weight. \emph{Centre: }Run time as a function of weight and number of particles in the IP regime. \emph{Right: }Run time as a function of the number of photons per MP in the IP regime, with a constant number of MPs $N_{MP}=100$.}}
\label{RunTimes}
\end{figure}

On a single workstation without any parallelisation, we conducted simulation runs in order to test the performance of the TrI LEns code. An overview of the run times in the IP regime for different weights and numbers of macro-particles can be seen in Figure \ref{RunTimes} in the central plot. We gradually increased both the number of MPs ($x$-axis) as well as the number of photons per MP (y-axis). The central plot shows, that a simulation with a small number of very "heavy" MPs performs better than a simulation with many, light MPs. For instance, a simulation with $100$ MPs, that correspond to $10^4$ photons each, terminates faster than a simulation with $10^5$ MPs, each consisting of $10$ photons. This is not very surprising, since the photons of each MP share the same energy and momentum and, therefore, are easier to handle than completely individual photons. At the same time, however, as the number of MPs grow, heavy MPs easily lead to a significantly worse performance, than simulation with even more, but lighter MPs. This can be discussed more accurately with a more detailed look at the run time as a function of the number of MPs with a fixed number of photons per MP (Figure \ref{RunTimes} left) and the opposite, a numbers of MPs and different numbers of photons per MP (Figure \ref{RunTimes} right). Here it becomes obvious, that the run time grows faster with number of photons per MP in the IP regime than it does with increased number of MPs. Nevertheless, in order to avoid extremely high numbers of MPs, increasing their weight can be very beneficial. The exact ratio between the number of MPs and the number of photons per MP has to be chosen carefully. For instance the central plot in Figure \ref{RunTimes} shows that a simulation with $10^4$ MPs, consisting of $10^3$ photons, ran significantly faster than a simulation with $10^3$ MPs with $10^4$ photons each. However, the opposite is true with $10^4$ MPs with $10^2$ photons per MP and $10^2$ MPs with $10^4$ photons each. Nevertheless, in general simulations with large number of MPs benefit from not too heavy MPs, while smaller simulations run faster with such heavy MPs. Please note, that the increase of the run-time in the central and right plot of Figure \ref{RunTimes} is in fact a feature of the IP representation. There, each of the photons of a MP is handled individually should a collision occur. In a PIC simulation or in the ST representation collisions of two MPs are handled in a statistical fashion and, therefore, the performance is independent of the weight of the MPs.\\
\newline
In order to increase performance multi-threading has been implemented. First tests showed no performance increase, however. The reason for this seems to be the lack of long loops of non-interacting instances, that would occur in grid-based simulations for instance. A straight-forward approach for parallelisation using multi-threading in the presented code leads to a significant over-head, that outweighs any performance gain. Less intuitive methods for multi-threading in tree codes exist, but have not been implemented, yet. Multi-core parallelisation still might also be a viable method of parallelisation. Spatial partitions of the hierarchies could be computed individually by separated cores.
\\
\newline

\subsection{Modelling the Breit-Wheeler process}
\label{BW}
\begin{figure}
\begin{minipage}[t]{0.45\textwidth}
	\centering
	\includegraphics[width = .7\textwidth, angle = 0]{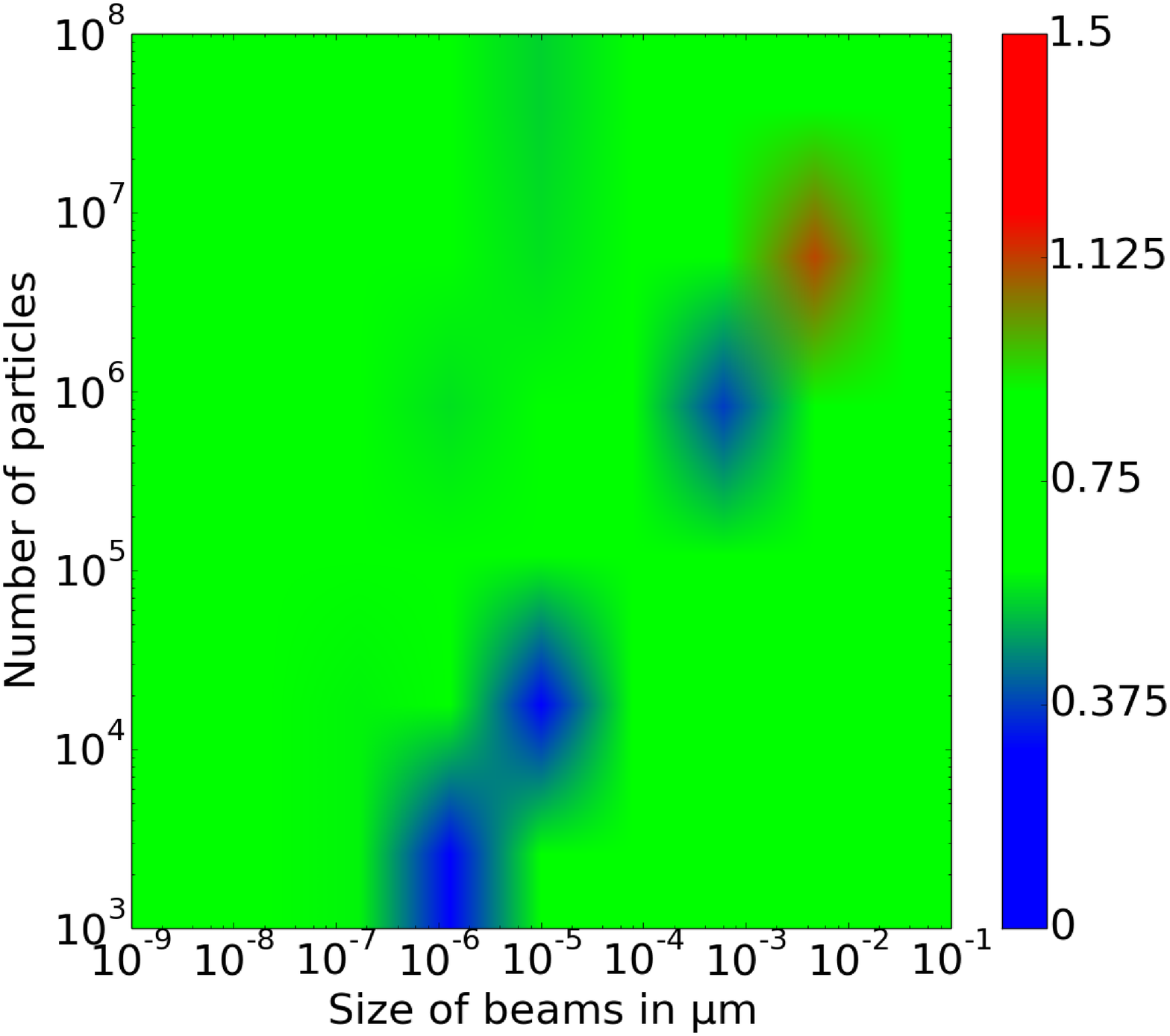}
\end{minipage}
\hspace{0.01\textwidth}
\begin{minipage}[t]{0.45\textwidth}
	\centering
	\includegraphics[width = .7\textwidth, angle = 0]{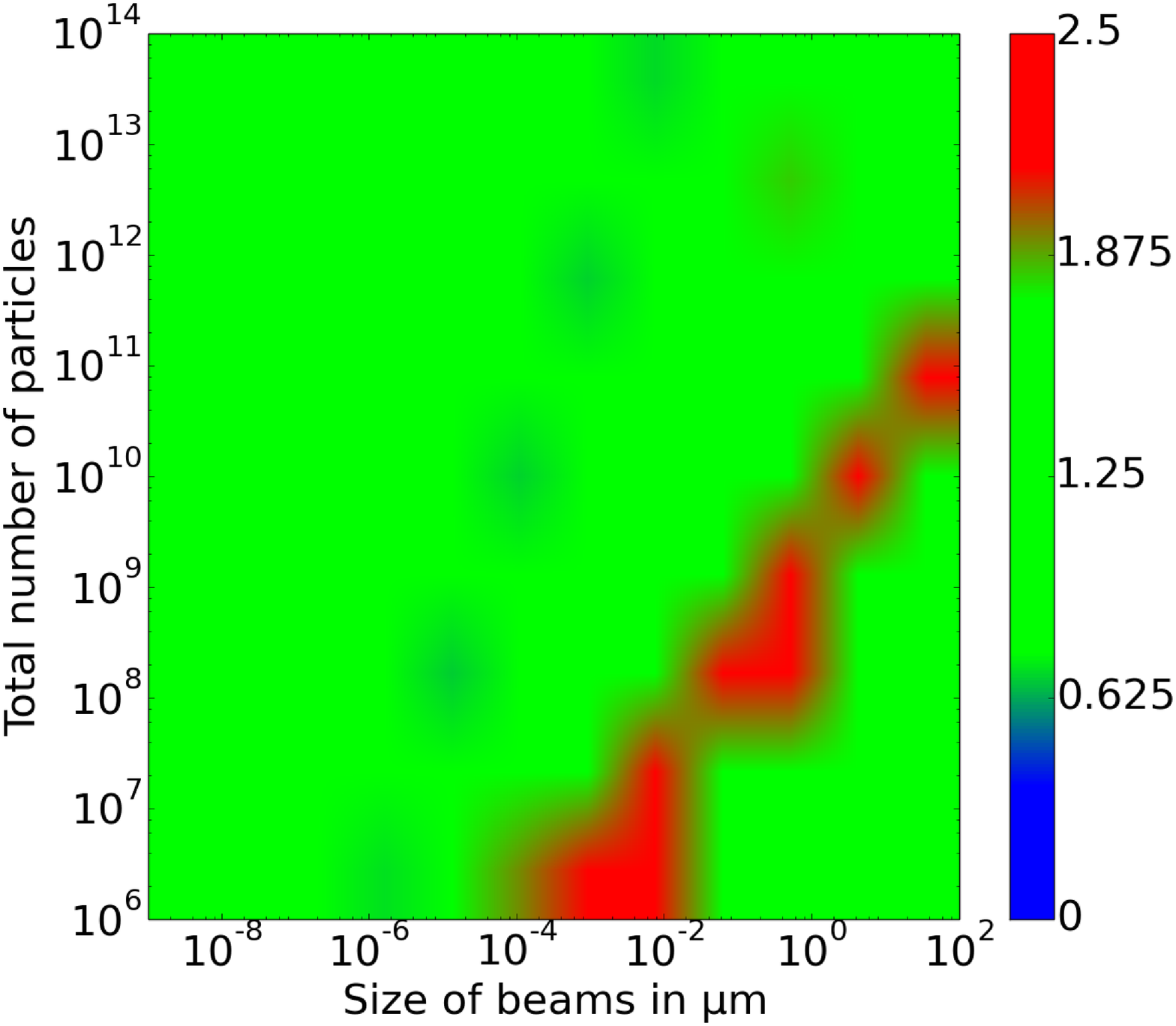}
\end{minipage}
\caption{\textit{Comparison of simulations with analytical predictions. Plotted is a ratio between simulation results and theoretical predictions of the number of created pairs as a function of the size of the beams and the total number of photons per beam. The beams consisted of $10^3$ MPs. \emph{Left: }Numbers of created pairs from IP simulations divided by analytical predictions. \emph{Right: }Results from ST simulations divided by analytical predictions.}}
\label{SimVsAna}
\end{figure}

As a physical application we chose to simulate the Breit-Wheeler (BW) process, described in section $1$. Here, we simulated two beams, consisting of $10^3$ MPs each, colliding with each other head-on ($\Phi_B = 180^\circ$). The number of photons per MPs and the total size of the beams were changed for different simulations. Each photon had a Lorentz-factor of $\gamma = 4$. The beams were placed at a distance of $200$ times their size from each other. All MPs received a random divergence from the main propagation direction, representing a total opening angle of the beam of $1^\circ$. We compared the resulting numbers of created pairs with analytical predictions for different beam sizes and number of photons. Since, the total number of created pairs depends strongly on the photon density and the simulation in the individual particle (IP) representation suffers from performance issues at very large total numbers of particles, those simulations were done with very small beams and interaction volumes. The simulations in the statistical (ST) regime were able to simulate realistic scenarios. Still, we decided to include some simulations with very small beams in order to have overlap, in order to compare both results directly. The left-hand side of Figure \ref{SimVsAna} shows the number of pairs created in the IP simulations $N_{Sim}$ divided by the theoretical predictions $N_{Theory}$, while the right-hand side shows the ratio of the results of the ST simulations and the analytical predictions. 

\begin{figure}
\begin{minipage}[t]{0.45\textwidth}
	\raggedleft
	\includegraphics[width = 0.8\textwidth, angle = 0]{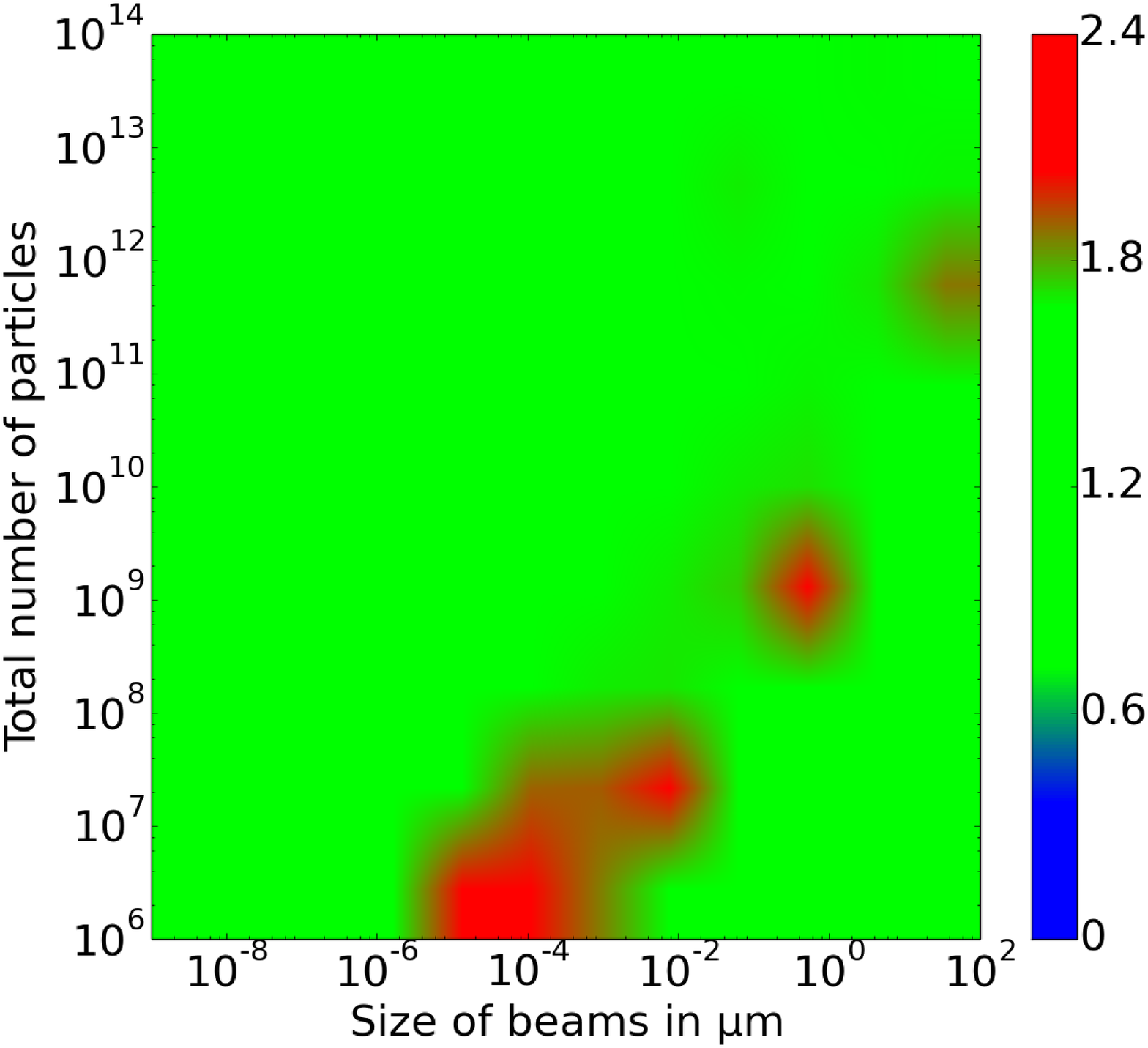}
\end{minipage}
\begin{minipage}[t]{0.45\textwidth}
	\raggedright
	\includegraphics[width = 0.8\textwidth, angle = 0]{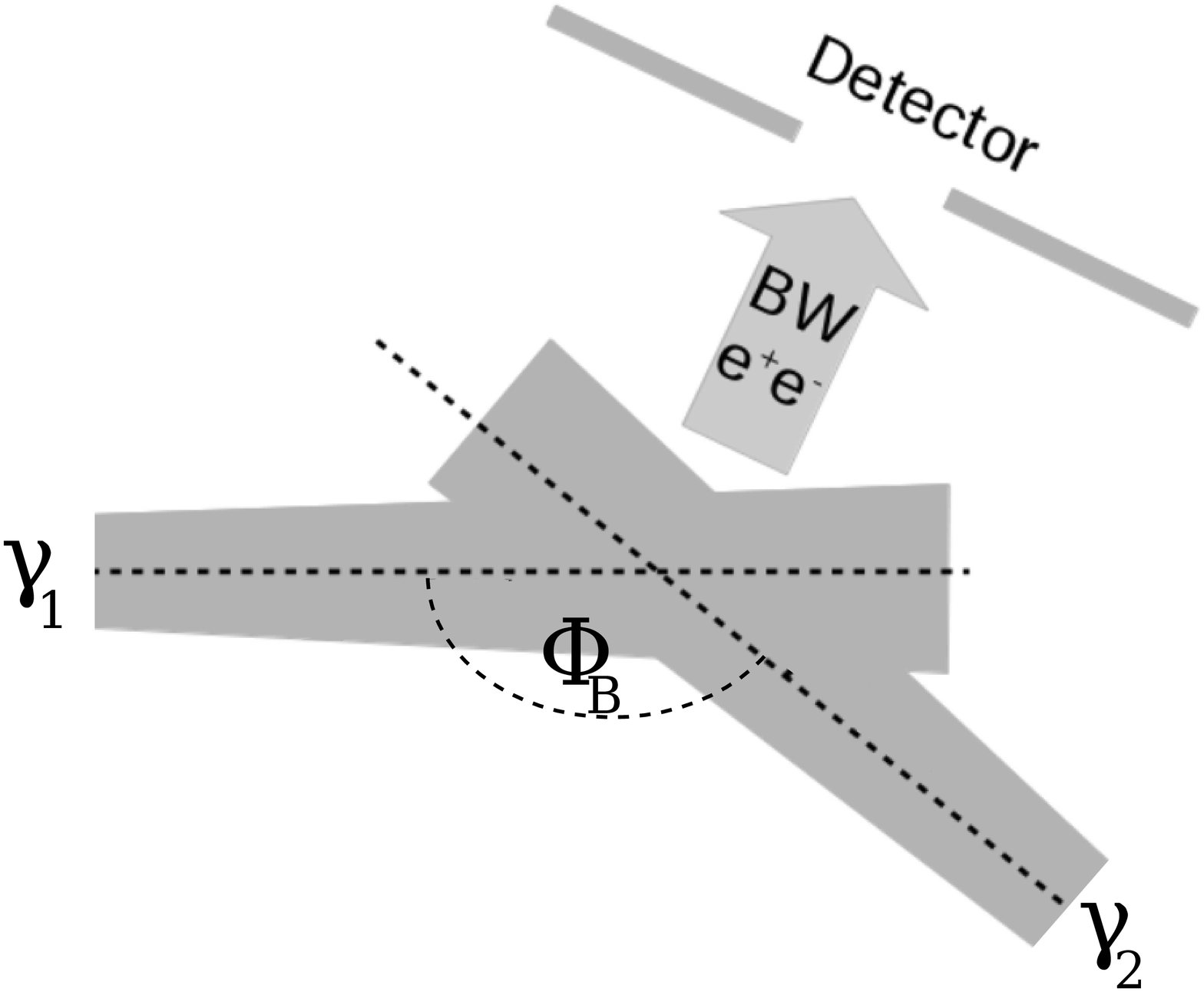}
\end{minipage}
\caption{\textit{\emph{Left: }Comparison of simulations of a BW-like scenario with analytical predictions similar to the right plot in Figure \ref{SimVsAna}, but with $10^4$ MPs in the ST representation. \emph{Right: }Proposed set-up for photon-photon collision. Two beams of $MeV$ photons collide in vacuum and create BW pairs.}}
\label{AddPic}
\end{figure}

The analytical results were calculated using (\ref{rate}) and (\ref{reactivity}) leading to
\begin{equation}
	\label{Ncollision}
	N_{\pm, \text{BW}} \sim n_{\gamma, 1} n_{\gamma, 2}V_\gamma \int_{\gamma_{1 \text{min}}}^{\gamma_{1 \text{max}}} \int_{\gamma_{2 \text{min}}}^{\gamma_{2 \text{max}}} f_{\gamma, 1}f_{\gamma, 2} \sigma_{\gamma\gamma} l_\gamma d\gamma_1 d\gamma_2,
\end{equation}
where $n_{\gamma,i}$ are the photon densities, $\gamma_i$ the photon energies and $f_{\gamma, i}$ are the distribution functions of the photons of the beam $i=1,2$, $V_\gamma$ is the volume of a beam, and $l_\gamma$ is its length. For details on (\ref{Ncollision}) see \cite{Xavier}.\\
 In both cases, green colour marks good agreement. Red colour stands for the case the simulations achieved higher numbers of collisions, while Blue means the opposite. It should be noted, that the colour bar is shifted, so that a perfect agreement is not necessarily displayed at a ratio of $r = N_{Sim} / N_{Theory} = 1$. The reason for this is, that even though, the parametric dependency of the number of created pairs on the size of the beams and the photons inside them, predicted by the theory, could be reproduced very well, the exact number of created pairs in the simulation differed slightly. The reasons for this are purely numerical, including a less than perfect sampling of phase space due to a limited number of MPs or rounding errors, while computing and handling the very small cross section. Since this deviation was the same for all simulations, simply shifting the colour-scale in Figure \ref{SimVsAna}, which is equivalent to multiplying a constant factor on the number of created pairs in every simulation, was enough to account for the difference. For comparison reason, we repeated the simulations, which lead to the right plot in Figure \ref{SimVsAna}, but with $10^4$ MPs. In the left plot in Figure \ref{AddPic} one can find, that the scaling factor decreased to a value closer to $1$, while the area left of the border shows an even better agreement with the theoretical predictions.\\
Both plots show rather large areas, where simulations and analytical predictions agree well. Each plot features a diagonal line, given by simulations, that produced a number of pairs, significantly different from the theoretical predictions. This line of simulations can be understood as a border between the area at small beams and high particle numbers and the simulations with large beams and low numbers of particles. The border separates two regimes. To its right, in the range of larger interaction volumes, simulations and theory perfectly agree in that, the photon density in this range is too low as to expect any collision at all. Since very small numbers of pairs are to be expected in the transition from this low-density regime, small deviations immediately lead to a significant differences from the theory. However, as soon as the number of expected pairs reaches larger values in the high-density regime to the left of the border, the statistics improve and simulations agree well with the predictions. 
We like to point out, that the two plots in Figure \ref{SimVsAna} are based on simulations using different particle representations (see section \ref{MPs} $4$). Due to the choice of parameters both plots have an overlap in the range of $10^6-10^8$ particles. In this range, the same qualitative behaviour can be found, despite the numerical differences, although, the red spot in the ST picture at $10^7$ particles and a volume size of $0.1\upmu m$ is more strongly developed in the IP simulations.\\
\newline
 

\begin{figure}
\begin{minipage}[t]{0.3\textwidth}
	\raggedleft
	\includegraphics[width = \textwidth, angle = 0]{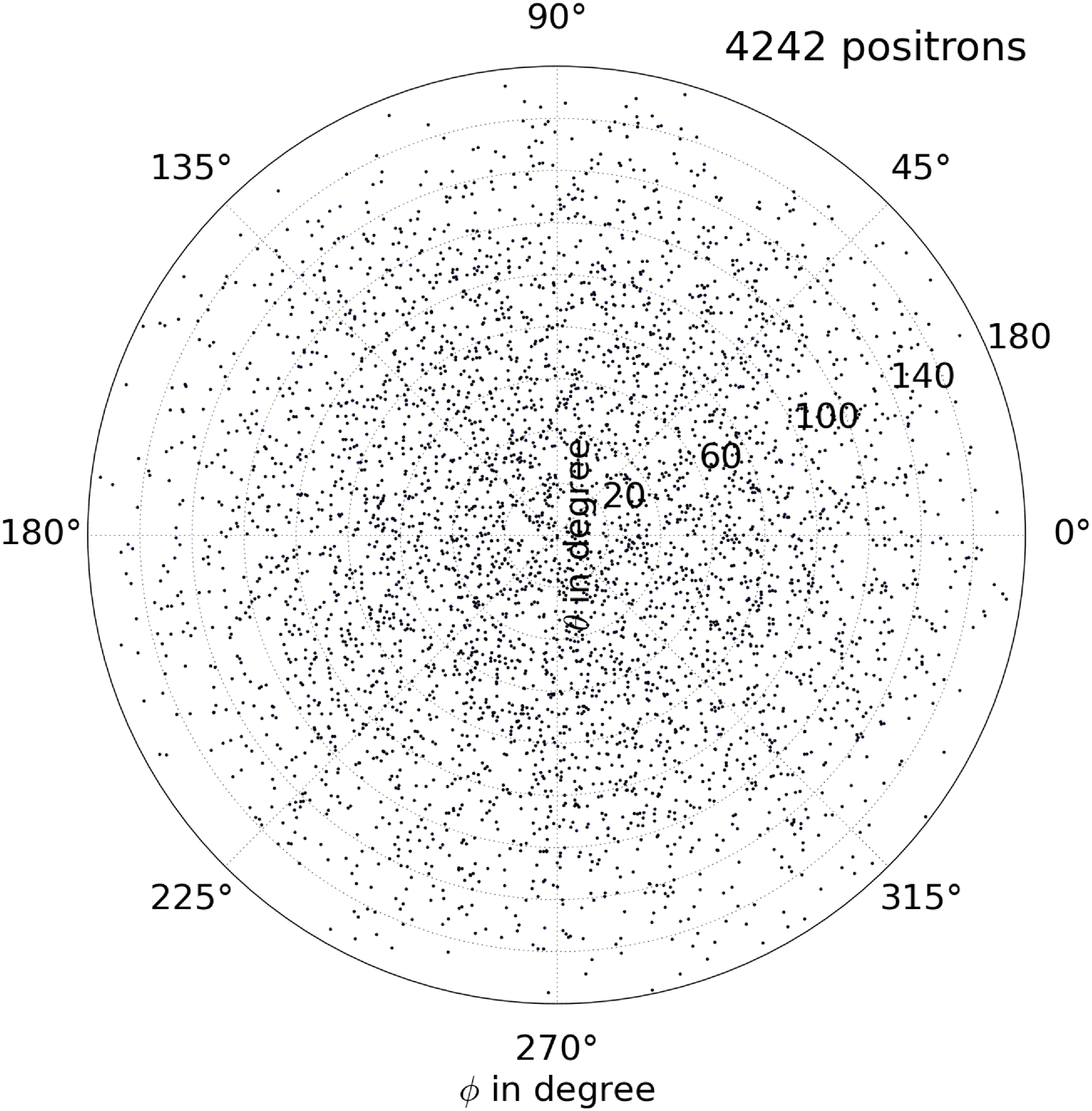}
\end{minipage}
\hspace{0.02\textwidth}
\begin{minipage}[t]{0.3\textwidth}
	\centering
	\includegraphics[width = \textwidth, angle = 0]{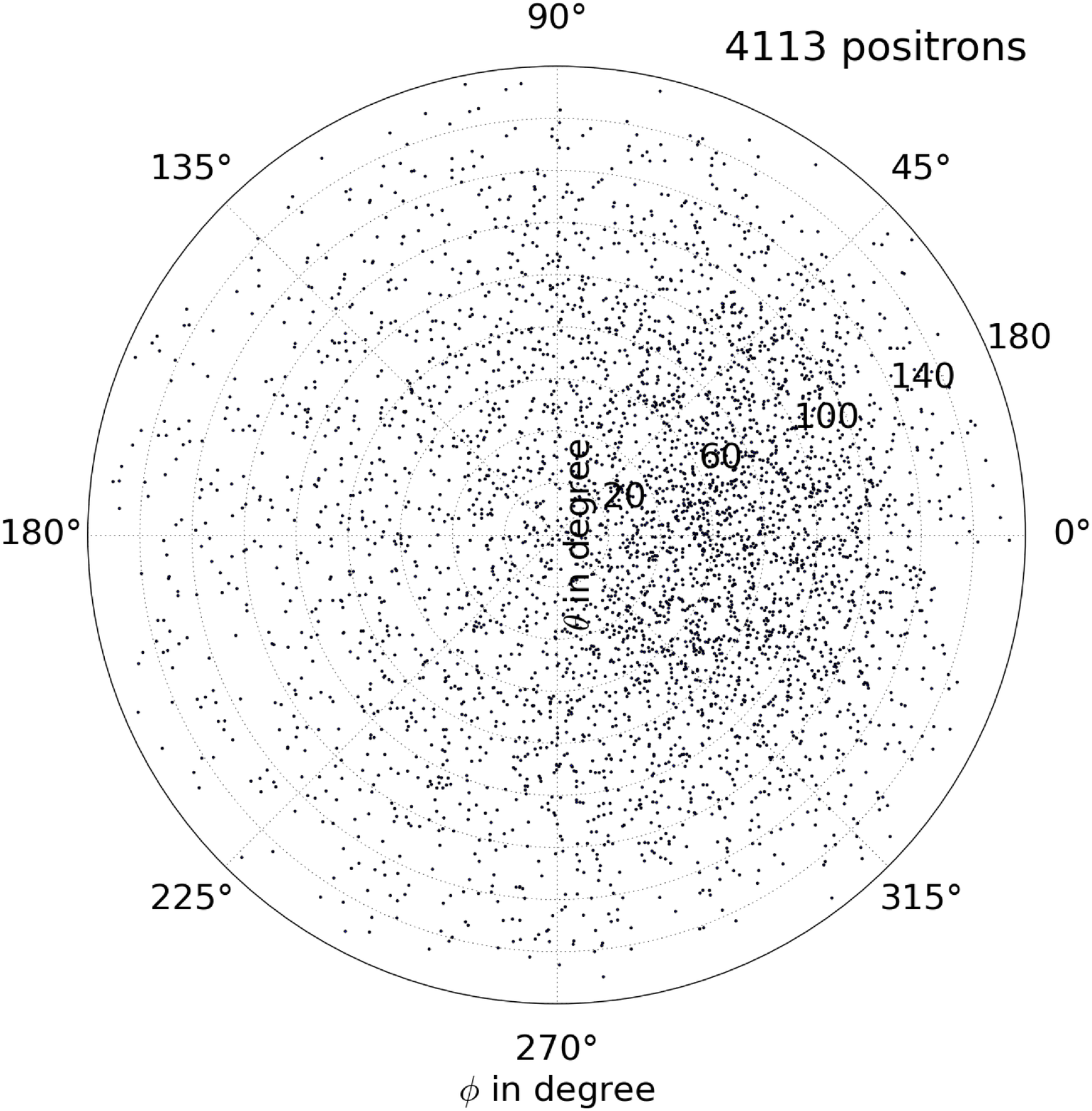}
\end{minipage}
\hspace{0.02\textwidth}
\begin{minipage}[t]{0.3\textwidth}
	\raggedright
	\includegraphics[width = \textwidth, angle = 0]{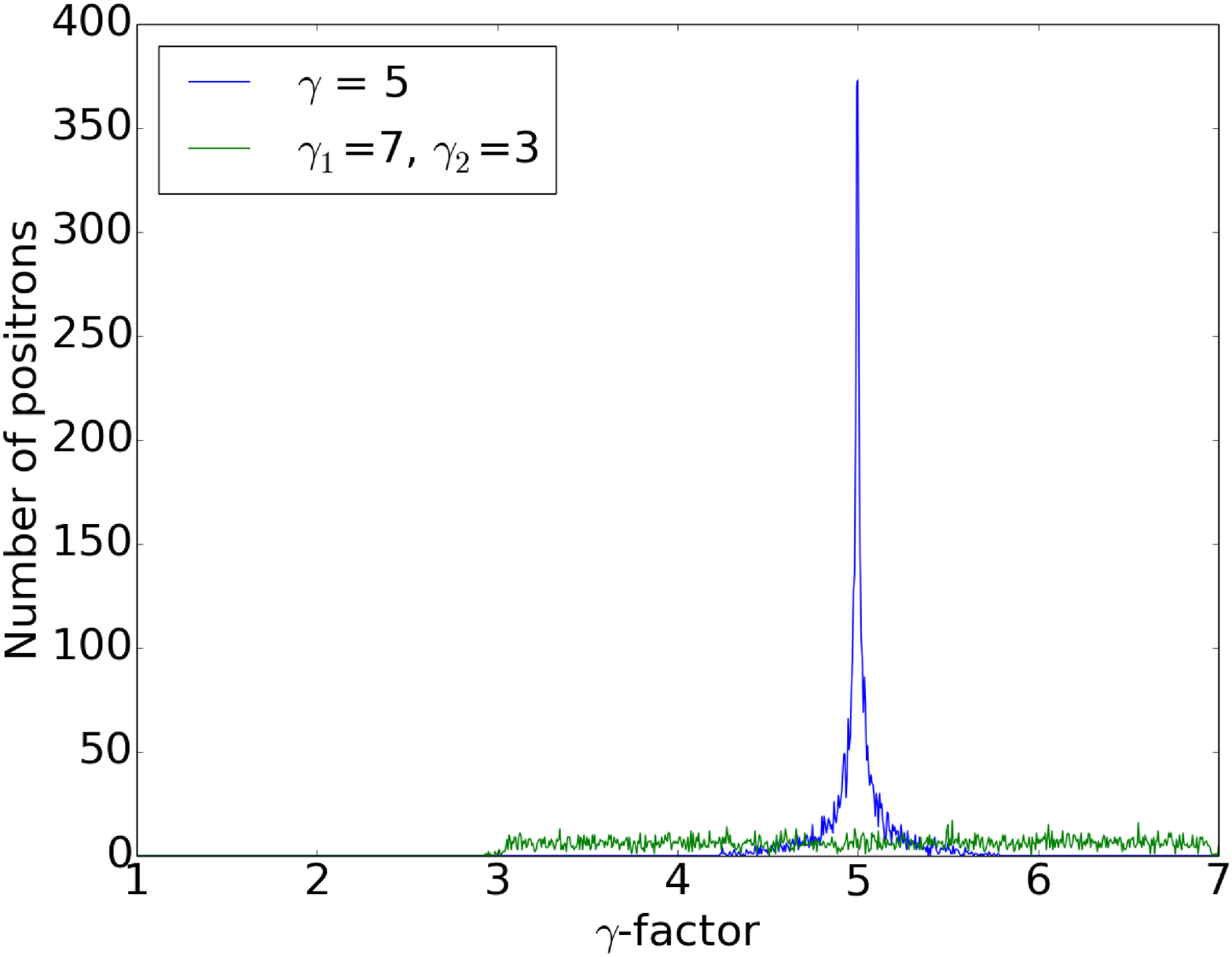}
\end{minipage}
\caption{Left: \textit{Momenta of pairs created by two beams (beam $1$ and $2$) of mono-energetic photons ($\gamma = 5$), simulated in ST scheme. Plotted is the direction of the momenta in spherical coordinates. The radius represents the polar angle. The beams are counter-propagating ($\Phi_B = 180^\circ$) , approaching from $\Phi_1 = 0^\circ$ and $\Phi_2 = 180^\circ$. \emph{Centre: }The same as on the left, but the Lorentz-factor of the photons of beam $1$ is $\gamma_1 = 7$ and of beam $2$ $\gamma_2 = 3$. \emph{Right: }Number of positrons as a function of their $\gamma$-factor for both simulations.}}
\label{Gamma}
\end{figure}

\begin{figure}
\begin{minipage}[t]{0.3\textwidth}
	\raggedleft
	\includegraphics[width = \textwidth, angle = 0]{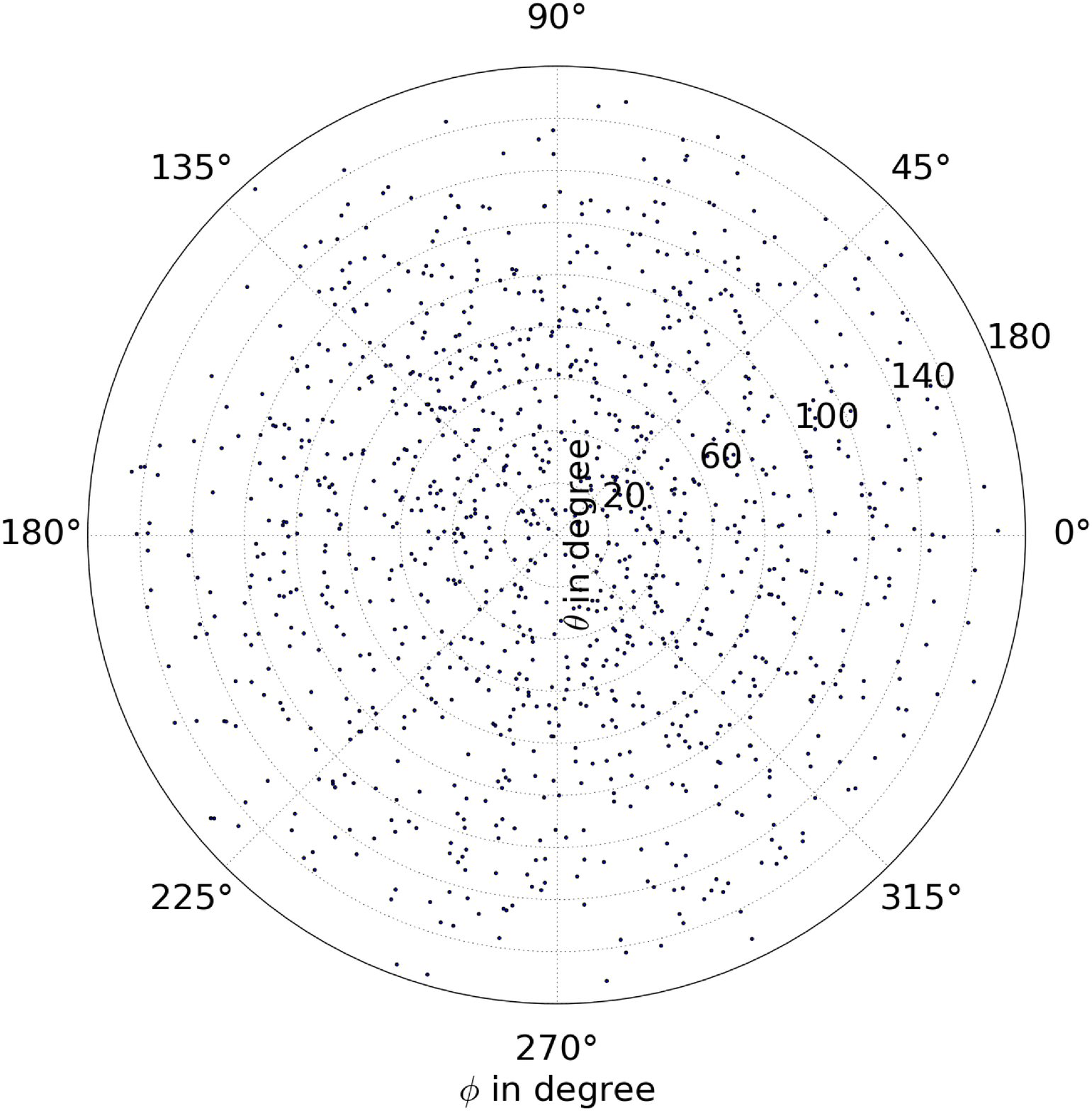}
\end{minipage}
\hspace{0.02\textwidth}
\begin{minipage}[t]{0.3\textwidth}
	\centering
	\includegraphics[width = \textwidth, angle = 0]{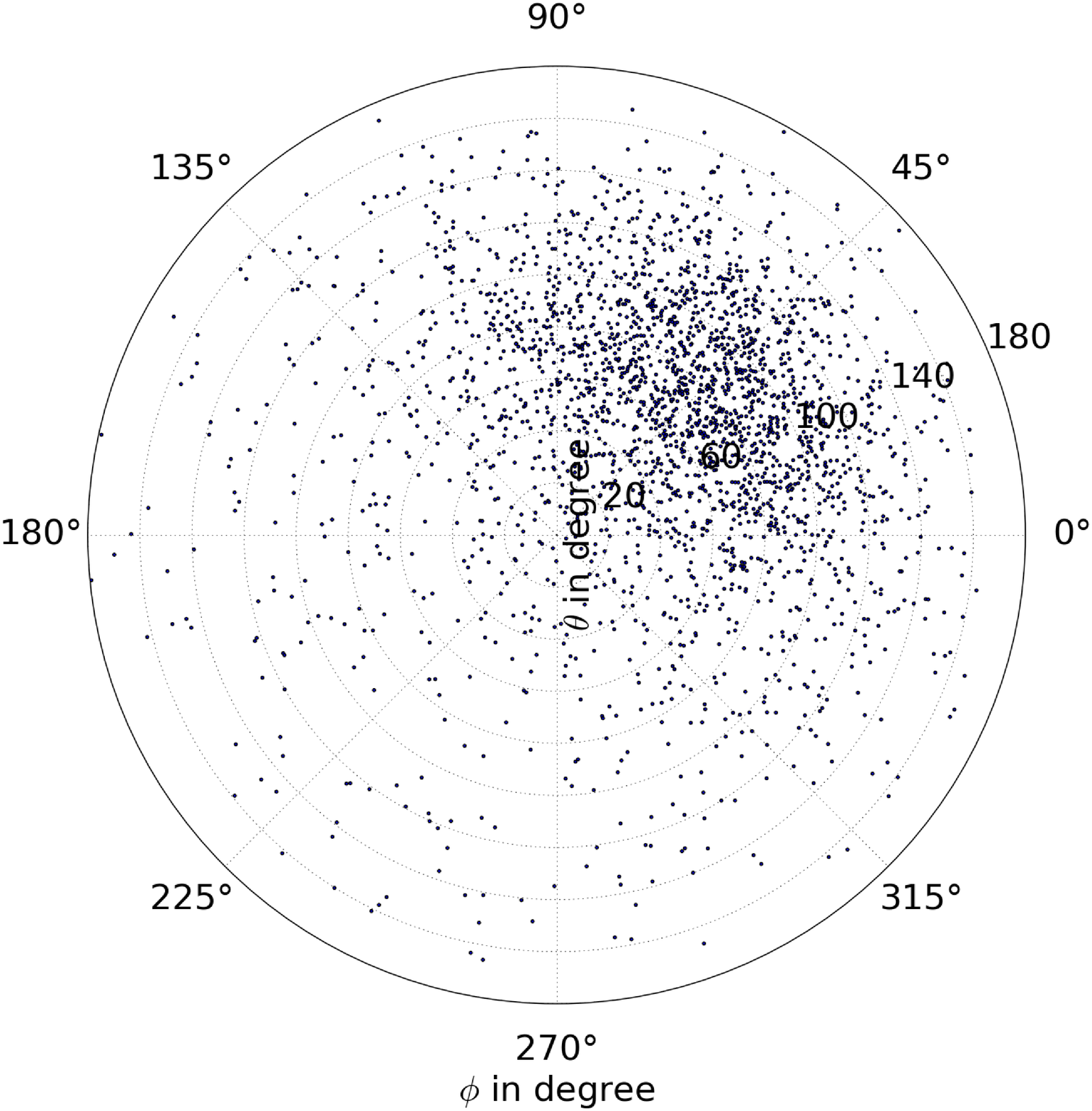}
\end{minipage}
\hspace{0.02\textwidth}
\begin{minipage}[t]{0.3\textwidth}
	\raggedright
	\includegraphics[width = \textwidth, angle = 0]{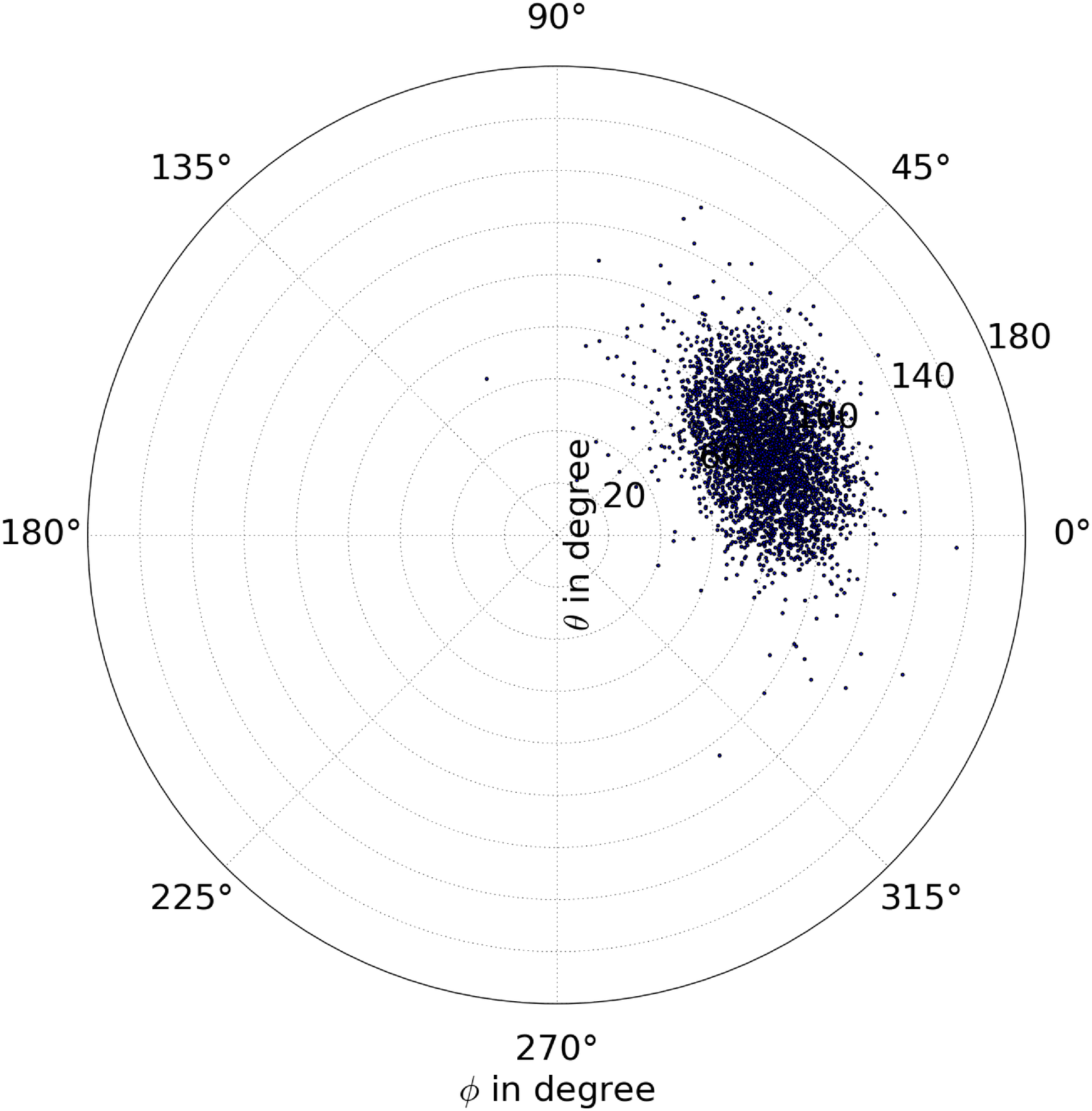}
\end{minipage}
	\caption{\textit{Angular distribution of positrons created by beams with a collision angle of \emph{left:} $\Phi_B = 180^\circ$, \emph{centre:} $\Phi_B = 90^\circ$ and \emph{right:} $\Phi_B = 45^\circ$. The polar angle of the direction of the momenta corresponds to the radius in the plot. One beam always has its origin at $180^\circ$, while the other one approaches from $0^\circ$ (left), $270\circ$ (centre) and $225^\circ$ (right).}}	
\label{Sims}
\end{figure}

In the following, we studied the impact of the angle $\Phi_B$ between the beams (right side of Figure \ref{AddPic}) and the photon energies on the momenta of created pairs. Using beams with different photon energies enabled us to change the distribution of created positrons. For the following simulations, we used two counter-propagating beams with a length and diameter of $4\upmu m$, consisting of $10^6$ MPs. The total number of photons was $2.5\cdot 10^{13}$, which meant, that we had to use simulations in the ST representation. Each MP had a random small transversal momentum, leading to a total opening angle of $20^\circ$ for each beam. The distance between both beams was $1$mm. In Figure \ref{Gamma} the angular distribution of momenta of positrons, created by two colliding beams is shown. Electrons are not represented, since they are created in a symmetric way. The plot shows the momentum of each MP, without taking into account, how many photons are represented by each one. Only the direction of each momentum is shown in spherical coordinates, where the radius in the plot corresponds to the polar angle of the direction vector, while the azimuthal coordinate represents the azimuthal part.  The plot on the left-hand side shows an isotropic distribution of momenta of positrons, created by two beams with an energy corresponding to a Lorentz-factor of $\gamma_1 = 5$ collide. In comparison to this, the central plot shows a distribution created in the collision of two beams, with photon energies corresponding to $\gamma_2=7$ and $\gamma_3=3$. The distribution clearly shifted towards the centre of mass (CoM) momentum, which points in the direction of propagation of beam $1$ according to (\ref{pTotal}). Figure \ref{Gamma} on the right-hand side shows the number of positrons as a function of their Lorentz-factor. The mono-energetic case shows a clear spike at the common Lorentz-factor, while the histogram of the case with $\gamma_2$ and $\gamma_3$ is almost homogeneously spread over the range of $3\le \gamma \le 7$. This can be understood by reviewing the transformation from the CoM-frame to the lab-frame. In the CoM-frame the momenta are spread homogeneously, since there is no preferred emission direction, if one ignores polarisation effects. For two beams with identical photon energy, the CoM momentum vanishes, making CoM-frame and lab frame identical and, thus, leading to the same homogeneous momentum distribution. However, the results are quite different in case of two beams with different photon energy and, therefore, a non-vanishing CoM momentum. Still, in the CoM-frame the momenta are distributed homogeneously. However, after applying the Lorentz-transformation, positron that are created with a momentum in the same direction as the CoM-momentum gain a higher momentum in the lab frame than positrons which in the CoM-frame have a momentum in the opposite direction. The width of the spread in the distribution of the Lorentz-factor of the positrons is given by (\ref{pTotal}). The energy that exceeds the threshold given by (\ref{eq.Cond}) is used to create this spread. This means, that two beams very close to the energy threshold would produce almost mono-energetic positrons. Both simulations ran on a single core of a work-station PC with $2\cdot 10^6$ macro-particles representing $5\cdot 10^{13}$ particles in total for less than six hours.\\
Returning to the BW experiment, proposed by Ribeyre \textit{et al.} \cite{Xavier}, we found, the set-up, that is shown in Figure \ref{AddPic} on the right, can be optimised by choosing an appropriate angle between the colliding beams in order to control the emission direction of the created pairs. In the following simulations, we applied similar conditions: two beams of $2.5\cdot 10^{13}$ photons with a Lorentz-factor of $\gamma=4$ and a total volume of $1\upmu$m$^3$ approach each other under different angles. We used $10^4$ MPs for each beam and the ST scheme. Figure \ref{Sims} shows the angular distribution of the momenta of created positrons, this time for different angles $\Phi_B$ between the two beams. Making use of the effect of the CoM momentum induced by angles other than $\Phi_B = 180^\circ$, the uniform angular distribution of pair momenta (left) is improved, so that more and more particles are emitted in a preferred direction (central and right). This provides us with a mean to create a more collimated beam of electron-positron pairs, making their detection and handling significantly easier.

\section{Conclusions}
Addressing the challenge of collision detections of a large number of particles we presented a novel tree-code TrI LEns, based on bounding volume hierarchies. Its main application at the time of this publication is the collision of photons according to the Breit-Wheeler process. We used our simulation code in order to investigate a proposed experiment on the detection of the Breit-Wheeler process, where $\sim 10^{13}$ photons collide with each other in a small volume of space. We showed, how the $N^2$ problem of collision detection, that would cause a significant challenge on most simulation codes, in particular PIC codes, can be mitigated by subdividing phase space. We compared the performance of the TrI LEns code to an equivalent mesh-based code and found that the tree code computed the problem faster than the mesh-based code by an order of magnitude, while being able to reproduce accurately analytical predictions for the Breit-Wheeler process. With small restrictions to the momentum distribution of particles, similar to the discretisation in PIC-codes, the performance can significantly be further improved.\\
Some of the possible further developments of this framework were already mentioned. Parallelisation by multi-threading, was implemented, but would require significantly more optimisation for achieving a performance gain. However, multi-processor parallelisation by partitioning space seems quite promising and would make the TrI LEns code more viable for computer-clusters.\\
Collision of arbitrary, charge-less particles like muons, neutrons, atoms, heavy particles and photons would require a new approach to the update method currently used for the tree hierarchy. However, even if a new hierarchy has to be computed every time step, we still expect a significant performance gain over conventional approaches.\\
Long range interactions, charged particles and external fields are not, yet, handled by the TrI LEns code. However, superposing another BV hierarchy specifically to long range interaction, similar to the BH-simulation, seems to be an opportunity for simulating arbitrary physical systems.
\section{Acknowledgements}
We acknowledge the financial support from the French National Research Agency (ANR) in the framework of "The Investments for the Future" programme IdEx Bordeaux - LAPHIA (ANR-10IDEX-03-02)-Project TULIMA. This work is partly supported by the Aquitaine Regional Council (project ARIEL).



\end{document}